\documentclass[preprint, eqsecnum]{emulateapj}

\usepackage{natbib}
\usepackage{graphicx}
\usepackage{amsmath}
\usepackage{color}

\newcommand{\PN}{\cal{P\!N\!}}

\slugcomment{Submitted to Astrophysical Journal}

\shorttitle{Post-Newtonian BBH Merger in Galactic Nuclei}
\shortauthors{Berentzen et al.}

\begin{document}

\title{Binary Black Hole Merger in Galactic Nuclei: Post-Newtonian
       Simulations}

\author{Ingo Berentzen\altaffilmark{1},
        Miguel Preto\altaffilmark{1},
        Peter Berczik\altaffilmark{1,2},
        David Merritt\altaffilmark{3} and
        Rainer Spurzem\altaffilmark{1}}

\altaffiltext{1}{Astronomisches Rechen-Institut, Zentrum f\"ur Astro\-no\-mie,
 Uni\-ver\-sity of Heidelberg, D-69120 Heidelberg, Germany}
\altaffiltext{2}{Main Astronomical Observatory, National Academy of Sciences
 of Ukraine, 27 Akademika Zabolotnoho St., 03680 Kyiv, Ukraine}
\altaffiltext{3}{Center for Computational Relativity and Gravitation, Ro\-ches\-ter
 Institute of Technology, 78 Lomb Memorial Drive, Rochester, NY 14623}
\email{iberent@ari.uni-heidelberg.de}

\begin{abstract}
 This paper studies the formation and evolution of binary supermassive
 black holes (SMBHs) in rotating galactic nuclei, focusing on the role
 of stellar dynamics. We present the first $N$-body simulations that
 follow the evolution of the SMBHs from kiloparsec separations all the
 way to their
 final relativistic coalescence, and that can robustly be scaled to
 real galaxies. The $N$-body code includes post-Newtonian ($\PN$\,)
 corrections to the binary equations of motion up to order $2.5$; we
 show that the evolution of the massive binary is only correctly reproduced
 if the conservative $1 \PN$ and $2 \PN$ terms are included. The orbital
 eccentricities of the massive binaries in our  simulations are often
 found to remain large until shortly before coalescence. This directly
 affects not only their orbital evolution rates, but has important
 consequences as well for the gravitational waveforms emitted
 during the relativistic inspiral.
 We estimate gravitational wave amplitudes when the frequencies fall
 inside the band of the (planned) Laser Interferometer Space Antennae (LISA).
 We find significant contributions --- well above the LISA sensitivity curve ---
 from the higher-order harmonics.
\end{abstract}


\section{Introduction}

 Supermassive black holes (SMBHs) are commonly observed at the centers
 of nearby galaxies, and the existence of quasars at redshifts $z \approx 6$
 implies that many of these SMBHs reached nearly their current masses at
 very early times. Bright galaxies are believed to form via hierarchical
 merging of smaller galaxies, and galaxy mergers lead inevitably to the
 formation of binary SMBHs (Begelman, Blandford \& Rees 1980). At their
 formation, such SMBH binaries typically have separations $a \approx a_{\mathrm h}$,
 where

\begin{equation}
 a_{\mathrm h} \equiv \frac{G \mu}{4 \sigma^2}
               \approx 3.3\,{\mathrm{pc}}~\frac{q}{(1+q)^2} \left( \frac{m_1+m_2}{10^8\,
               {\mathrm M}_\odot} \right)^{0.6}\  ,
\end{equation}

 \noindent
 \citep[see, e.g.,][]{MM05}.  This ``hard binary separation''
 $a_{\mathrm h}$ is defined in the standard way as the value of the binary
 semi-major axis at which passing stars are ejected with
 velocities high enough to unbind them from the SMBHs; $a_{\mathrm h}$
 is a function of the binary reduced mass $\mu \equiv m_1\,m_2/(m_1+m_2)$
 and the ambient stellar velocity dispersion $\sigma$ ($q \equiv m_2/m_1 \le 1$).
 At least one binary SMBH has probably been observed, with a projected
 separation of about $7$\,pc \citep{Rod06}. The quasi-periodic outbursts
 of the quasar OJ287 have also plausibly been modeled as a binary system
 with separation $\sim 0.05$ pc and eccentricity $\sim 0.6$ \citep{Val07}.
 A few other galactic systems are known to contain dual SMBHs with
 separations of a few kiloparsec, presumably the precursors of binary
 SMBHs \citep{kom2003,bia08}.

 Binary SMBHs may eventually coalesce, but only after stellar- and/or
 gas-dynamical processes have first brought the two SMBHs to separations
 small enough ($\sim 10^{-3}$\,pc) that gravitational wave emission is
 effective.  Whether Nature typically succeeds in overcoming this ``final
 parsec problem,'' or whether long-lived binary SMBHs are the norm, is currently
 unknown. Persistence of the binaries would have a number of potentially
 important consequences, including ejection of SMBHs from galaxy nuclei via
 three-body interactions and a reduction in the mean ratio of SMBH mass
 to galaxy mass \citep[Volonteri, Haardt \& Madau 2003;][]{Madau04,Libes06}.
 In addition, plans to detect gravitational waves from the final inspiral
 of binary SMBHs by space-based observatories such the Laser Interferometer
 Space Antennae (LISA) (Sesana, Volonteri \& Haardt 2007) might need to be
 re-considered if evolution of the
 binaries typically stalls at large separations.

 In a spherical, gas-free galaxy, the evolution of a binary SMBH slows down  at separations of $\sim a_{\mathrm h}$ because stars on orbits that
 intersect the
 binary -- so-called ``loss-cone orbits'' -- are depleted in just a few
 galaxy crossing times (e.g., Begelman et al. 1980, Makino et al. 1993,
 Makino 1998, Berczik, Spurzem \& Merritt 2005; Merritt 2006). A massive
 binary can continue to evolve in such a galaxy only if the loss-cone
 orbits are somehow repopulated. Gravitational scattering is one possible
 mechanism for loss-cone
 repopulation, but it is only effective in low-luminosity galaxies
 with central relaxation times below $\sim 10$\,Gyr \citep{Yu02, MMS07}.
 Dense concentrations of gas can substantially accelerate the evolution
 of a massive binary by increasing the drag on the individual SMBHs
 \citep{Escala04,Escala05,Dotti07}. The plausibility of such gas
 accumulations, with masses comparable to the masses of the SMBHs,
 is unclear however, particularly in the most massive galaxies. Galaxy
 merger simulations including gas have followed the two SMBHs until
 separations of a few tens of parsecs, roughly the hard binary separation
 \citep{Kazant05,Mayer07}. These simulations contain useful information
 about the formation and early evolution of SMBH binaries but -- at
 least so far -- do not have the resolution needed to address the final
 parsec problem.

 An alternative pathway exists for binary SMBHs to evolve beyond
 $a \approx a_{\mathrm h}$, even in gas-free galaxies with long relaxation
 times.  If the galaxy potential is non-axisymmetric, many of the stars
 will be on centrophilic orbits, i.e., orbits that pass near the center of
 the galaxy once per orbital period \citep{GB85}. If even a small
 fraction of a galaxy's mass is associated with such orbits, the
 ``feeding'' rate of a central binary can be enormously enhanced
 compared with the rates in spherical galaxies \citep{MPoon04}.
 \citet{BMSB06} explored this pathway in $N$-body simulations containing particle
 numbers up to one million; the initial galaxy models were rotating,
 and in some cases the rotation was rapid enough to induce the
 formation of a triaxial bar. Evolution of the SMBH binary was observed
 not to stall at $a \approx a_{\mathrm h}$ in the triaxial models;
 furthermore the evolution rates exhibited no discernible dependence
 on particle number, as predicted if the mechanism of loss-cone
 refilling is collisionless. To date, the \citet{BMSB06} simulations
 are the only ones that have successfully followed the evolution of
 a binary SMBH all the way from galactic to sub-parsec scales, and
 that can robustly be scaled to real galaxies due to the lack of an
 appreciable $N$-dependence of the evolution rates.

 The final binary separation in the \citet{BMSB06} simulations was
 $\sim 0.05 a_{\mathrm h}$, just small enough that gravitational wave
 emission would induce coalescence in $10$\,Gyr (assuming a scaling
 that assigns a mass $10^8 {\mathrm M}_\odot$ to the binary SMBH). In this
 paper, we repeat the Berczik et al. simulations using a new
 $N$-body code that incorporates the post-Newtonian ($\PN$~) corrections
 to the equations of motion of the SMBHs. This allows us to follow
 the evolution of the binary all the way to final coalescence. In
 addition, we are able to estimate the strength of the gravitational
 waves emitted during the final inspiral. A key result is that the
 eccentricity of the binary can remain high until shortly before coalescence.
 Both the strength of the emitted gravitational waves, and the
 timescale for coalescence, are strongly dependent on the binary
 eccentricity. Highly eccentric black hole binaries would represent
 appropriate candidates for forthcoming verification of gravitational
 radiation through the planned mission of LISA.

 The outline of this paper is as follows: in Section~2 we give a description
 of the numerical methods and initial models used in this work. In Section~3
 we then present the results of a set of $N$-body simulations using both the
 classical, i.e., Newtonian gravity (Sec.~3.1) and the relativistic, i.e.,
 post-Newtonian gravity (Sec.~3.2). In Section~4 we discuss
 our results in a broader astrophysical context, focusing on the SMBH merger
 as sources of gravitational wave emission. In Section~5 we finish with the
 conclusions.

\section{Numerical modeling}

 For the $N$-body simulations presented in this work we use a modified
 version of the publicly
 available\footnote{http://wiki.cs.rit.edu/view/GRAPEcluster/phiGRAPE}
 $\varphi$\,-{\sc Grape} code \citep{HGMS07}. This code is based on an
 {\sc Nbody1}-like algorithm \citep{Aar99}, including a hierarchical
 timestep scheme and a $4^{\mathrm{th}}$-order Hermite integrator \citep[see, e.g.,][]{MA92}.
 The gravitational acceleration $\mathbf{a}$ and its first time-derivative
 ${d \mathbf{a}}/{d t}$ (also called: jerk)  between the {\em field}
 particles -- representing the 'stellar' component in the galactic
 nucleus -- are calculated using the special-purpose hardware {\sc Grape}-6A
 (Fukushige, Makino \& Kawai 2005). We also use the {\sc Grape} hardware
 to calculate the interaction between the field particles and the black holes
 (and vice versa). The acceleration between the two black hole (BH) particles,
 the corresponding jerk as well as the relativistic corrections, are all
 calculated using the hosts CPU.

 To determine the timestep bin for each individual particle we use the
 standard criterium \citep{Aar85} of the form:

 \begin{equation}
 \Delta t_i = \sqrt{\,\eta\ \frac{|{\bf a}_i|\,|{\bf a}_i^{(2)}| +
               |{\dot{\bf a}_i}|^2}{|{\dot{\bf a}_i}|\,|{\bf a}_i^{(3)}| +
               |{\bf a}_i^{(2)}|^2} } \ ,
 \label{Eq1}
 \end{equation}

 \noindent where ${\bf a}_i, {\dot{\bf a}}_i$, and ${\bf a}_i^{(k)}$ are
 the acceleration, its first and $k^{\mathrm{th}}$ time-derivatives, respectively,
 of a given particle $i$. For details on how to obtain the higher-order
 time-derivatives we refer to \cite{HGMS07}.  Whereas we use
 $\eta = \eta_{\star}\!=\!2 \times 10^{-2}$ for the field particles, 
 we use a smaller $\eta_{\mathrm{BH}}= 0.1\,\eta_{\star} = 2\times10^{-3}$ for
 the two BH particles in order to guarantee an adequate conservation of
 energy and momentum (in the Newtonian gravity simulations). Moreover,
 the two BHs are always advanced synchronously using the smaller
 $\Delta t_{\mathrm{BH}}$ of the two.

 The simulations have been carried out on the dedicated high-performance
 GRAPE-6A clusters at the
 Astronomisches Rechen-Institut in Hei\-del\-berg,\footnote{GRACE: see http://www.ari.uni-heidelberg.de/grace}
 at the Ro\-chester Institute of Technoloy,\footnote{gravitySimulator: see http://www.cs.rit.edu/~grapecluster}
 and at the Main Astronomical Observatory in Kiev.\footnote{golowood:  http://www.mao.kiev.ua/golowood/eng}

\subsection{Initial Conditions} \label{SecInit}

 The initial conditions of the $N$-body models used in this work are
 chosen to be similar to the ones used by Berczik et al. \citeyearpar{BMS05}
 and \citet{BMSB06}, allowing for a direct comparison to their simulations
 and for an accordant interpretation of our results. Here, we briefly describe
 the main aspects of the initial model set-up and refer the reader to the
 latter two publications (and references therein) for further details.

 The initial field particle distribution -- representing the galactic
 nucleus -- is set-up following the phase-space density distribution of a
 King model \citep{King66} with internal rotation \citep[e.g.,][and
 references therein]{LL96, EGFJS07}. The models have been set up with a
 central concentration parameter of $W_0=6$ and an initial rotation of
 $\omega_0\equiv\sqrt{9/(4\pi\,G\,\rho_{\mathrm c})}\,\Omega_0=1.8$, where
 $G$ and $\rho_{\mathrm c}$ are the gravitational constant and the central
 concentration, respectively. The net rotation in our models is chosen
 to be counter-clockwise, with the angular momentum vector being aligned
 with the $z$-axis. We adopt the standard $N$-body units
 \citep[see, e.g.,][and references therein]{Aar03b} to our numerical
 models by setting both the gravitational constant $G$ and the
 total mass $M_{\star}$ of the $N$-body system to unity, and by
 setting its total energy to $E_{\star}=-1/4$.

 In this work the galactic nucleus is represented with total numbers of
 $N_{\star}=25\times10^3$ and $50\times10^3$ field particles, respectively,
 each with nine different random realizations. We restrict ourselves to
 these relatively small particle numbers as a trade-off for the large set
 of simulations that we have performed in grand total. However, as
 \citet{BMSB06} have shown, the hardening rate of SMBH binaries in such
 rotating galaxy models is in this regime essentially independent of the
 field particle number $N_{\star}$.
 {\color{black} The main difference between our models
 and the latter ones is the $\PN$ treatment of the two SMBHs. Therefore
 the results of our simulations are expected to be quasi-$N$-independent
 as well, since we use a rotation parameter of $\omega_0 > 1.2$ \citep{BMSB06}.}
 The field particles have all equal mass $m_{\star} = M_{\star}/N_{\star}$.
 The gravitational softening length for the field particles has been set
 to $\epsilon_{\star}=10^{-4}$ (in model units). The same softening is
 also used for the star$\leftrightarrow$BH interaction. The two BH
 particles themselves are evolved without any gravitational softening,
 i.e., with $\epsilon_{\mathrm{BH}}=0.0$.

 We set the masses $m_1$ and $m_2$ of the two BH particles to be one
 per cent of the nuclei mass $M_{\star}$, i.e., in our case $m_1=m_2=0.01$.
 The two BH particles are initially placed on the $y$-axis at $y_{1,2}=\pm0.3$,
 respectively, within the $z=0$ plane. The BHs are given an initial velocity
 $v_x$, corresponding to roughly ten per cent of the local circular velocity,
 as derived
 from the enclosed mass of the underlying field particle distribution. With
 this we get roughly $v_{1,2}\approx\pm0.07$, in our model units. Note that in this
 configuration the two black holes are initially unbound with respect
 to each other.

 To scale our $N$-body results to real galaxies we consider the
 total energy of the system

\begin{equation}
  E = - \alpha \, \frac{G\,M^2}{R} ,
\end{equation}

\noindent
 where $G$, $M$ are the gravitational constant and the total
 mass of our model system (typically some fraction of the bulge 
 and cusp components in a galactic nucleus), and $R$, $\alpha$ are
 model dependent quantities, giving a scaling radius and a numerical
 constant of order unity, respectively. As an example, for a non-rotating
 King model with $W_0=6$ (which is used in our models as an initial
 configuration) we would have $R=r_{\mathrm c}$ as the core radius
 as defined by \citet{King66} and $\alpha=0.0759$. 

 With $G=1$ we can freely choose two of the three scales (energy, radius,
 or mass). In standard $N$-body units the unit of mass is the total mass, i.e.
 $M=1$, and the unit of energy is $4\,E$ ($E=-0.25$). Therefore the radial
 unit is determined by the condition $R = 4\,\alpha$ (this value determines
 the value of $R$ in $N$-body units, e.g., for a Plummer model we have
 $\alpha = 3\pi / 64$ and $R = 3\pi /16$). The physical units of mass, length,
 energy, velocity and time are then given as:

\begin{eqnarray}
 \left[ M \right] & = & M                                     \nonumber \\
 \left[ L \right] & = & \frac{R}{4\alpha}                     \nonumber \\
 \left[ E \right] & = & 4\alpha \, \frac{G\,M^2}{R}           \nonumber \\
 \left[ V \right] & = & 2\sqrt{\alpha} \left(\frac{G\,M}{R}\right)^{1/2} \nonumber \\
 \left[ T \right] & = & \frac{1}{8\,\alpha^{3/2}} \left(\frac{R^3}{G\,M}\right)^{1/2}  \ .
\end{eqnarray}

\noindent
The speed of light $c$ in $N$-body units is then

\begin{eqnarray}
 c & = & c_0 / \left[ V \right] = c_0 \cdot \left( 4\alpha \frac{G\,M}{R} \right)^{-1/2}
\nonumber \\
   & = & 457  \left( {\frac{M}{10^{11}\,{\mathrm M}_{\odot}}} \right)^{-1/2}
          \left( {\frac{R}{10^3\, {\mathrm{pc}}}} \right)^{1/2} \ ,
\end{eqnarray}

\noindent
 where $c_0$ is the speed of light in physical units, and for the
 second expression we have inserted $\alpha=-0.25$. This illustrates
 that in the post-Newtonian regime the $N$-body problem is not
 scale-free anymore - to fix the scale for $c$ means fixing the radial scale or
 vice versa. In our simulations we have chosen different values for the
 speed of light, i.e., $c=447, 141, 44,$ and $14$, in order to enhance
 the effect of the relativistic corrections. At the same time, this
 variation of $c$ scales the Schwarzschild radius
 $R_{\mathrm{BH}}=2\,G\,M_{\mathrm{BH}}/c^2$ of the two BHs to values
 of $10^{-7}$, $10^{-6}$, $10^{-5}$ and $10^{-4}$, respectively, in model
 units.
 
\subsection{Relativistic treatment of compact binary systems}

 Super-massive black hole binaries are expected to be subject to the emission
 of gravitational waves. Since the gravitational waves drain energy and angular
 momentum from the binary, its orbital elements will change in the course of
 their dynamical evolution. Already more than four decades ago \cite{PM63} and
 \cite{P64} derived the change of energy $E$ and of the orbital elements for
 a Keplerian orbit, namely its semi-major axis $a$ and eccentricity $e$, under
 the influence of the gravitational wave (quadrupole) emission. The orbit averaged
 expressions for two compact objects with masses $m_1$ and $m_2$ orbiting each
 other are given as:

\begin{eqnarray}
\label{dadtpm}
 & & \left\langle\!\frac{da}{dt}\!\right\rangle  =
 -\frac{64}{5} \frac{G^3\, m_1\, m_2 (m_1+m_2)}{c^5\, a^3} f(e), \\
 & & \left\langle\!\frac{de}{dt}\!\right\rangle  =
  -\frac{304}{15} \frac{G^3\, m_1\, m_2 (m_1+m_2)}{c^5\,a^4
  (1-e^2)^{5/2}} \left(e+\frac{121}{304}e^3\right),  \\
 & & \left\langle\!\frac{dE}{dt}\!\right\rangle  =
  -\frac{32}{5} \frac{G^4\, m_1^2\, m_2^2 (m_1+m_2)}{c^5\,a^5} f(e) , 
\label{dedtpm}
\end{eqnarray}

 \noindent where the so-called enhancement factor $f(e)$, given by the
 following function, shows a strong dependence on the orbital eccentricity
 $e$:

\begin{equation}
 f(e) = \left(1-e^{2}\right)^{-7/2}\left(1+\frac{73}{24}e^{2}+\frac{37}{96}e^{4}\right) .
\label{enhance}
\end{equation}

 \noindent We note that the rate at which energy $E$ is lost due
 to the emission of gravitational waves depends strongly on the
 eccentricity $e$ of the orbit. Hence, compact binaries with high
 eccentricities (i.e., small $a$ and $e\approx1$) are expected to be
 strong sources of gravitational waves. The equations of Peters \& Mathews
 describe the shrinking of the semi-major axis (corresponding to an
 inspiral of the two objects) and the circularization of the orbit.
 According to \cite{PM63} and \cite{P64} the typical timescale of
 coalescence due to the emission of gravitational radiation is given by

\begin{equation}
  t_{\mathrm{gr}} = \frac{5}{64} \frac{c^{5}a_{\mathrm{gr}}^{4}}{G^{3}m_{1}\,m_{2}(m_{1}+m_{2})\,f(e)} \ ,
\label{Eq.tgr}
\end{equation}

 \noindent
 where $a_{\mathrm{gr}}$ denotes the characteristic separation for gravitational
 wave emission.

 To account for the relativistic effects in our numerical $N$-body
 simulations we use the post-Newtonian formalism. The equations of
 motion for a compact binary system are written in harmonic
 coordinates \citep{Sch85} and are
 defined in the inertial $N$-body frame of reference. As they are relativistic,
 they are  (1) invariant under $\PN$-expanded Lorentz transformations, (2) reduce
 to the geodesics of the $\PN$-expanded Schwarzschild metric in the limit
 in which one of the masses goes to zero and neglecting the spin, and (3) are
 conservative if the $2.5 \PN$~radiation reaction term is turned-off. In fact,
 an isolated $2 \PN$ binary should conserve its generalized $\PN$ integrals of
 motion up to ${\cal O}(1/{\mathrm c}^6)$ \citep{ABF01}. 
 We implement the $\PN$ equations
 of motion formulated in the absolute Euclidean space and time of Newton
 \cite[e.g.,][equation~168 therein]{Bla06}. In the present work we apply all
 $\PN$ corrections up to the order ${\cal O}(1/{\mathrm c}^5)$, i.e., the
 $2.5 \PN$ corrections is the highest order that we take into account.
 While the $2.5 \PN$ correction accounts for the emission of gravitational
 waves, the $1 \PN$ and $2 \PN$ terms conserve energy. However, the latter
 two terms result in precession of the orbital pericenter. Similar to the
 equations of motion in the center of mass frame (Kupi, Amaro-Seoane \& Spurzem 2006),
 one can write the accelerations, e.g., for particle $1$ of the binary,
 in the following form:

 \begin{equation}
 {\bf a}_1 =  - \frac{G\,m_2}{r_{12}^2} \left[ (1+{\cal A})\,
 {\bf n}_{12} + {\cal B}\,{\bf v}_{12} \right]\,,
 \end{equation}

 \noindent where $r_{12}$ is the separation of the two particles,
 ${\bf n}_{12}$ the normalized relative position vector, and
 ${\bf v}_{12}$ is the relative velocity. The two functions
 $\cal{A}$ and $\cal{B}$ contain the different orders of the
 $\PN$ expansion and can be written as:

 \begin{eqnarray}
  {\cal A} & = & \frac{1}{c^2}\, {\cal{A}}_{1{\cal P\!N}} +
                 \frac{1}{c^4}\, {\cal{A}}_{2{\cal P\!N}} +
                 \frac{1}{c^5}\, {\cal{A}}_{2.5{\cal P\!N}} +
                 {\cal O}\left( \frac{1}{c^6} \right) \, \\
  {\cal B} & = & \frac{1}{c^2}\, {\cal{B}}_{1{\cal P\!N}} +
                 \frac{1}{c^4}\, {\cal{B}}_{2{\cal P\!N}} +
                 \frac{1}{c^5}\, {\cal{B}}_{2.5{\cal P\!N}} +
                 {\cal O}\left( \frac{1}{c^6} \right) .
 \end{eqnarray}

 \noindent In this notation, the first post-Newtonian correction
 ($1 \PN$) is, for example, given as:

 \begin{eqnarray}
  {\cal A}_{1{\cal P\!N}} & = & \left[ \frac{5 G\,m_1}{r_{12}} +
  \frac{4 G\,m_2}{r_{12}} + \frac{3}{2}({\bf n}_{12}{\bf \cdot}{\bf v}_2)^2 - {\bf
v}_1^2 + \right.  \nonumber \\
  & + & \left. \hspace{2ex} 4({\bf v}_1{\bf \cdot} {\bf v}_2) - 2{\bf v}_2^2
         \right] \, , \\
   {\cal B}_{1{\cal P\!N}}  & = & 4\,({\bf n}_{12}{\bf \cdot} {\bf v}_1) -
                                  3\,({\bf n}_{12}{\bf \cdot}{\bf v}_2) \, .
 \end{eqnarray}

 \noindent
 We forgo from wri\-ting down the higher order $\PN$ cor\-rec\-tions
 due to their lengthy form -- especially the one of the $2 \PN$ correction.
 The complete expressions are given in, e.g., \citet{Bla06} and
  \citet{KASS06}.

 Our numerical implementation of the $\PN$ corrections to the accelerations
 -- and particularly to the corresponding jerk as required for the Hermite
 integration scheme -- has been thoroughly tested against other available
 codes, such as the ones used by \citet{KASS06} and \citet{Aar07}. A
 detailed description of the numerical tests and the results
 are given in \citep{BPBMS08}.
 {\color{black} An alternative implementation specially tailored to the
 treatment of single massive BH within galactic nuclei can be found
 in \citet{LB08}.}
 Our implementation of the $\PN$ corrections allows us to turn on and off
 the different $\PN$ orders separately at will. In the current set of
 models presented here we apply the selected $\PN$ terms to the two
 BH particles at all times during the simulations.

 Within the $\PN$ framework, the energy lost due to gravitational wave
 emission is given by \citep[see][equation~171 therein]{Bla06}:

\begin{eqnarray}
 \frac{dE}{dt} = \frac{4}{5} \frac{G^2 m_1^2 m_2}{c^5 r_{12}^3}
 \left[({\bf v}_1 {\bf v}_{12}) \left(-{\bf v}_{12}^2 + 2\frac{G m_1}{r_{12}} -
 8 \frac{G m_2}{r_{12}} \right) \right. & + & \nonumber \\
  + \left. ({\bf n}_{12} {\bf \cdot} {\bf v}_1)({\bf n}_{12}{\bf \cdot}{\bf v}_{12}) \left(3 {\bf v}_{12}^2
 - 6\frac{G m_1}{r_{12}}+\frac{52}{3}\frac{G m_2}{r_12}\right)\right] & + &
\nonumber  \\
  + \ (1 \leftrightarrow 2) + {\cal{O}}(c^{-7}). \hspace{4.8cm} 
\label{dedtpn}
\end{eqnarray} 

\noindent Note that this latter equation gives -- in contrast to the
 orbit averaged expression in Eq.~\ref{dedtpm} -- the {\em instantaneous}
 energy loss due to the emission of gravitational waves. We will apply
 both equations, i.e.,  Eq.~\ref{dedtpm} and Eq.~\ref{dedtpn},
 for comparison in our analysis.

\section{Results}

\subsection{Purely Newtonian simulations - the fiducial case} \label{SecN}

\begin{figure*}[ht]
\begin{center}
 \includegraphics[width=1.0\textwidth]{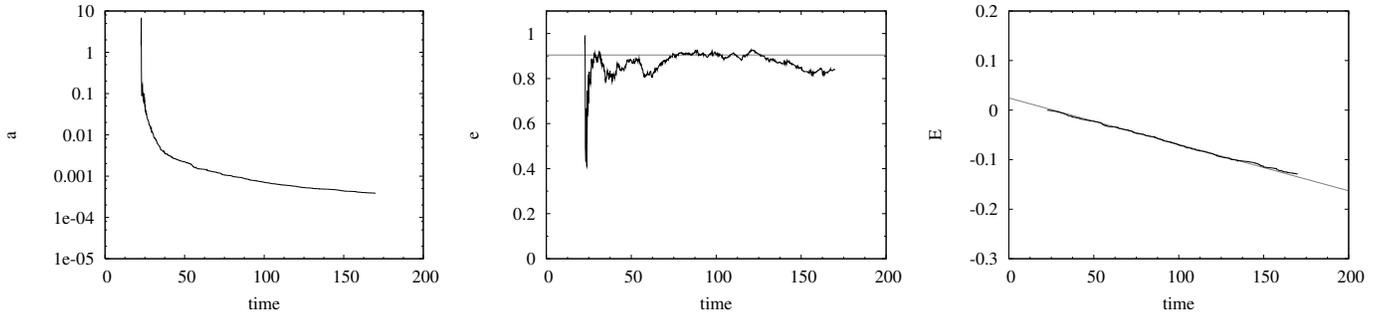}
\end{center}
\caption{Example of the typical evolution of the SMBH binaries
 semi-major axis $a$ (left panel), its eccentricity $e$ (middle
 panel) and its total energy $E$ (right panel) as a function of time
 in the classical (Newtonian) gravity models. The lines show the
 average eccentricity after the binary has formed and the linear
 fit to the energy loss due to super-elastic scattering. All
 quantities are given in model units.}
\label{fig1}
\end{figure*}

%
\begin{figure}
\begin{center}
 \includegraphics[width=1.0\columnwidth]{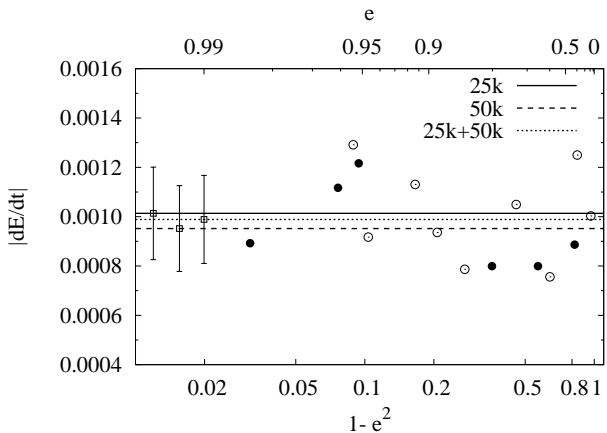}
\end{center}
\caption{ 
 {\color{black}
 Average rate of energy loss due to purely Newtonian,
 stellar-dynamical effects as a function of the binaries eccentricity.
 The symbols represent models with $25$k (open circles) and $50$k
 (filled circles) field particles, respectively. The horizontal
 lines indicate the mean values of $|dE/dt|$ for our set of $25k$
 (full line) and $50k$ (dashed line) models. The mean value of the
 combined set is plotted with a dotted line.  Note that in our
 models the energy loss, or equivalently the binaries hardening
 rate, is basically independent of the eccentricity and -- in contrast to
 spherically symmetric nuclei models -- {\color{black} also essentially $N$-independent
 within the indicated error-bars.}}
}
\label{fig2}
\end{figure}

 In this section we describe the results of our purely Newtonian
 simulations, i.e., classical models without any $\PN$ corrections.
 For these fiducial models, the evolution of the binary BHs and
 the field particles is very similar to the corresponding models
 reported by \cite{BMSB06}. Due to the relatively high degree of
 rotation in the models used in this work, the initially
 axisymmetric nucleus models become dynamically unstable and rapidly
 develop a rotating triaxial (bar-like) structure. As a result
 of this global instability and due to the dynamical friction
 against the stellar background, the two BHs are quickly
 funneled towards the density center of the nucleus, where they
 form a binary system in our models typically after some
 $\Delta t \approx 20$, or some $30$\,Myr.

 It actually turns out that certain binaries characteristics,
 such as the time when the binary forms ($t_{\mathrm{form}}$), as well
 as its orbital elements, are very sensitive to the initial
 conditions of the stellar distribution. Since the eccentricity
 of the binary is a crucial quantity for its subsequent evolution
 towards relativistic inspiral and coalescence, it is important to
 obtain a statistical sample.  Therefore we decided to use nine
 different realizations of the same galaxy nucleus model, by
 sampling the underlying distribution function with different
 initial random seeds. This way, e.g., we find eccentricities
 of the binaries typically in the range between $0.4$ up to $0.99$.
 The binaries in our simulations tend to form with the higher
 eccentricities around $0.9$. However, it cannot be decided
 at this point whether stochastic encounters between the BHs and the stars,
 or the global (non-linear) bar-instability plays the dominant
 role in determining the final binary properties.

%
\begin{figure}[t]
\begin{center}
 \includegraphics[width=1.0\columnwidth]{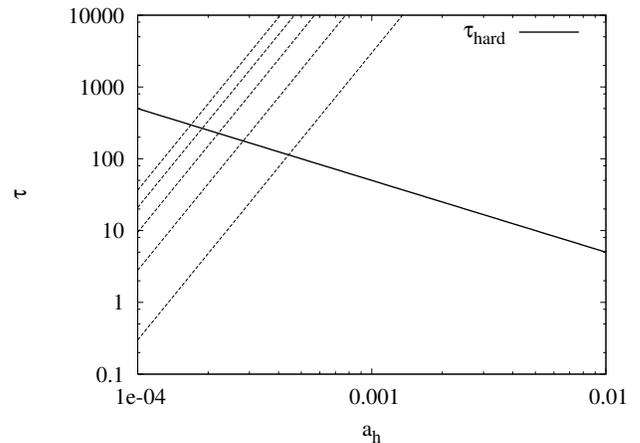}
\end{center}
\caption{Characteristic timescales as a function of the semi-major
 axis $a_{\mathrm h}$ for a) the Newtonian hardening (thick full line)
 and b) Peters \& Mathews (dotted lines) for orbits with
 eccentricities $e$ ranging from $0.5$ to $0.9$ (from top to bottom)
 in steps of $\Delta e=0.1$. The speed of light has been chosen as
 $c\approx500$ in model units.}
\label{figT}
\end{figure}

%
\begin{figure*}
\begin{center}
 \includegraphics[width=1.0\textwidth]{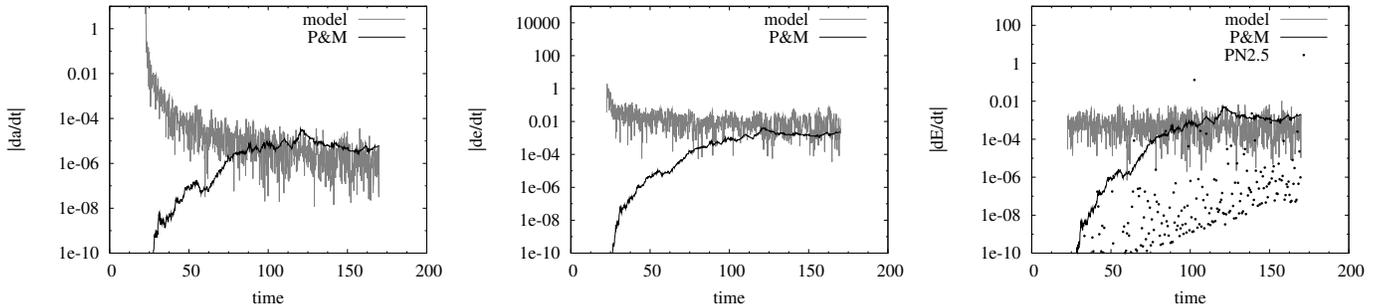}
\end{center}
\caption{Time-derivatives of the orbital elements $a$, $e$ and $E$ (from left
 to right). The gray curves show the discrete differences as calculated from
 the model directly, and the black curves show the derivatives based on 
 the Peters \& Mathews formalism. The dotted curve in the right panel
 shows the results from the post-Newtonian expression (see Eq.~\ref{dedtpn} in
 the text).}
\label{fig4}
\end{figure*}

 In Fig.~\ref{fig1} we show an example for the typical time
 evolution of the (Keplerian) orbital elements for one of
 the SMBH binaries in our $50$k Newtonian models. In the panels, from
 left to right, we show the evolution of the semi-major axis $a$,
 the eccentricity $e$, and the energy $E$ of the binary, respectively.
 Note that the extremely large values of $a$ and $e$ at early times
 are due to the fact the the two SMBH are still unbound. Only
 when they have formed a gravitationally bound pair, i.e., when the
 binaries energy $E$ becomes negative, the eccentricity $e$ remains
 below a value of $1$. As one can see in the middle panel,
 the eccentricity varies only mildly in the subsequent evolution,
 i.e., when the binary is hard and interacts with the surrounding
 field particles mainly by super-elastic three-body scattering.
 Therefore, we calculate some mean eccentricity $\bar{e}$ for each
 model as indicated by the horizontal line in the middle panel of
 Fig.~\ref{fig1}. We find $\bar{e}$ to be a useful quantity to
 characterize the binary, but it is important to bear in mind
 that it provides only a first order approximation, since the
 real eccentricity may in fact vary with time due to stochastic
 encounters with field particles.

 We find that the loss of energy (Fig.~\ref{fig1}, right panel) 
 due to super-elastic three-body encounters is almost constant
 after roughly $t\gtrsim30$. The straight line shows the result
 of a linear least-squares fit to $E(t)$ after the binary has
 formed. In Fig.~\ref{fig2} we show the resulting slopes $|dE/dt|$
 as derived from such linear fitting as a function of $(1-e^2)$ for
 our different Newtonian models.\footnote{This quantity ($1-e^2$)
 is used rather then $e$ itself, because it enters with a power
 of $-7/2$ in the enhancement factor f(e) (see Eq.~\ref{enhance}
 in the text).} The horizontal lines in Fig.~\ref{fig2} indicate
 the mean values of $dE/dt$ for both the separate and combined
 sets of our Newtonian $25$k and $50$k models. 

 Since $E \propto (1/a)$ for a Keplerian orbit, the rate $dE/dt$ provides
 a direct measure for the hardening rate of the binary $d/dt\,(1/a)$.
 We find that the hardening rate shows an essentially $N$-independence of both
 the eccentricity of the binaries' orbit and the number of field particles
 used in the models.

 The latter result is in good agreement with \cite{BMSB06} who found
 only small variations in the hardening rate for models with particle
 numbers ranging from $25$k to $1$M field particles, which is interpreted
 as a result of an efficient loss-cone refilling as found in such triaxial,
 rotating nuclei models.

 {\color{black}
 For our two Newtonian sets of $25$k and $50$k models, we find mean values
 for $|dE/dt|$ of about $(1.01 \pm 0.19) \times 10^{-3}$ and
                        $(0.95 \pm 0.17) \times 10^{-3}$, respectively.
 The hardening rate in our models is found to be essentially $N$-independent
 within the given error margins. This finding is clearly in accord with the
 results of \cite{BMSB06}. The mean value of $dE/dt$ for the combined set
 of models thus results in being roughly $\dot{E}\equiv dE/dt \approx-0.001$.
 }

 One can use the results of the previous paragraph to make some simple
 estimates regarding the binary evolution timescale in the purely
 Newtonian regime described above. For a Keplerian orbit we get:

\begin{equation}
   \frac{da}{dt} = \frac{2\,a^2}{G\,m_{1}\,m_{2}} \frac{dE}{dt}  \, .
\label{Eq4}
\end{equation}

 \noindent We define the binary hardening time in the usual way as

\begin{eqnarray}
 \tau_{\mathrm{hard}} & \equiv & \left|{\frac{1}{a}}{\frac{da}{dt}}\right|^{-1}
= \left|{\frac{1}{E}}{\frac{dE}{dt}}\right|^{-1} \nonumber \\
 & = & {\frac{G\,m_1\,m_2}{2a}} \left|{\frac{dE}{dt}}\right|^{-1}.
\end{eqnarray}

 \noindent
 Writing the rate of change of the binary energy in $N$-body units
 as $|dE/dt|=10^{-3}\times K$, where $K\approx 1$, this becomes in
 physical units

\begin{eqnarray}
&& \tau_{\mathrm{hard}} = 9.80 \times 10^3 K^{-1} {\frac{m_1\,m_2\,r_{c}^{5/2}}{
 {G}^{1/2} M_{\mathrm{gal}}^{5/2} a}} \nonumber \\
&&\approx 4.62\times 10^4 {\mathrm{yr}} {\frac{m_1\,m_2}{\left(10^8\,{\mathrm M}_\odot\right)^2a}}
\left({\frac{r_{c}}{10^2 {\mathrm{pc}}}}\right)^{5/2}
\left({\frac{M_{\mathrm{gal}}}{10^{11}M_\odot}}\right)^{-5/2}
\left({\frac{a}{1\ {\mathrm{pc}}}}\right)^{-1}
\end{eqnarray}

 \noindent
 This expression shows clearly that a constant rate of energy
 loss corresponds to a gradually increasing timescale for binary 
 hardening.
 If we assume that the expression holds for arbitrarily small $a$,
 we can predict a time to coalescence of

\begin{eqnarray}
t_{\mathrm{coal}} - t_0 = 
\tau_{\mathrm{hard}}(a_0)\left({\frac{a_0}{a_{\mathrm{coal}}}} - 1\right) ,
\end{eqnarray}

\noindent
 where $a_0=a(t_0)$ and $a_{\mathrm{coal}}=a(t_{\mathrm{coal}})$
 \citep[][]{FF05}.
 Setting $a_0=1$\,pc and $a_{\mathrm{coal}}=(10^{-2} \ldots 10^{-4})\,a_0$ 
 gives coalescence times ranging from $\sim 10$ Myr to $1$ Gyr for
 a galaxy with $M_{\mathrm{gal}}=10^{11} {\mathrm M}_\odot$ and $r_{\mathrm c}=100$\,pc.
 Thus we see clearly that the binaries in our galaxy models would have no
 difficulty reaching very small separations in a Hubble time, even
 in the absence of relativistic energy loss.

%
\begin{figure*}[ht!!!]
\begin{center}
 \includegraphics[width=1.0\textwidth]{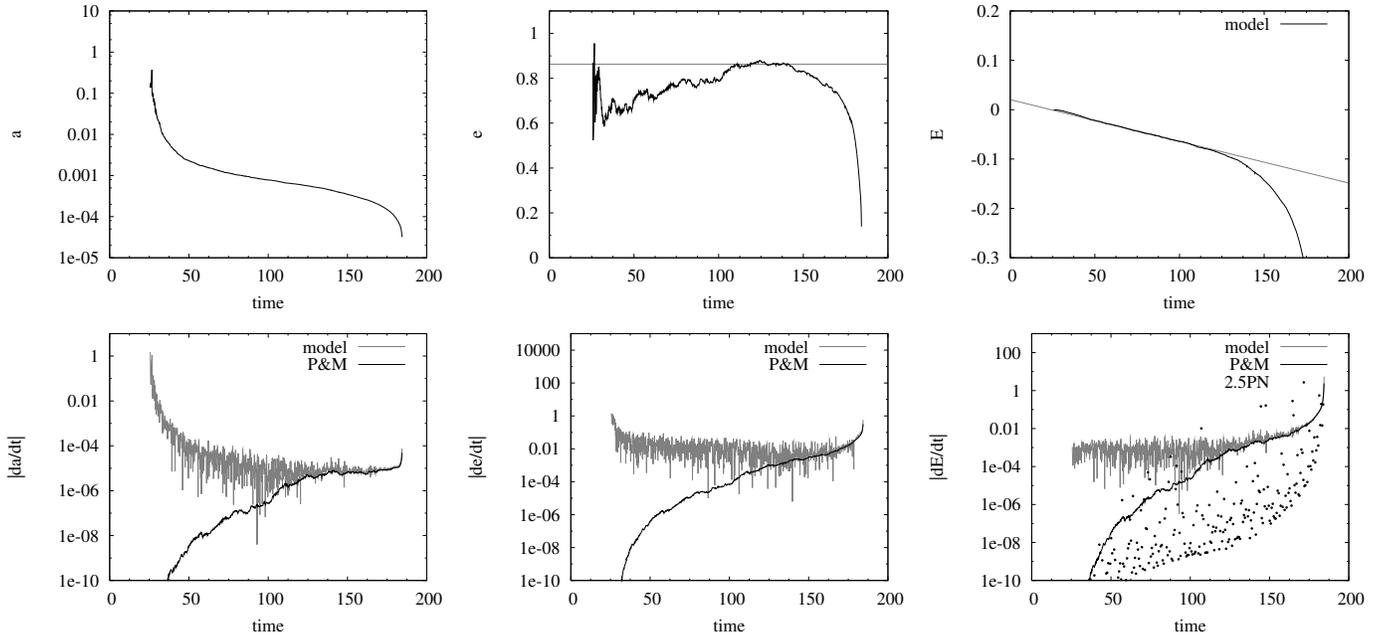}
\end{center}
\caption{Time evolution of a SMBH binary orbit in a Newtonian + $2.5 \PN$ model
 (using $c=447$). Layout for the upper and lower panels are the same as in
 Figs.~\ref{fig1} and \ref{fig4}, respectively.}
\label{fig5}
\end{figure*}

%
\begin{figure*}[ht]
\begin{center}
 \includegraphics[width=1.0\textwidth]{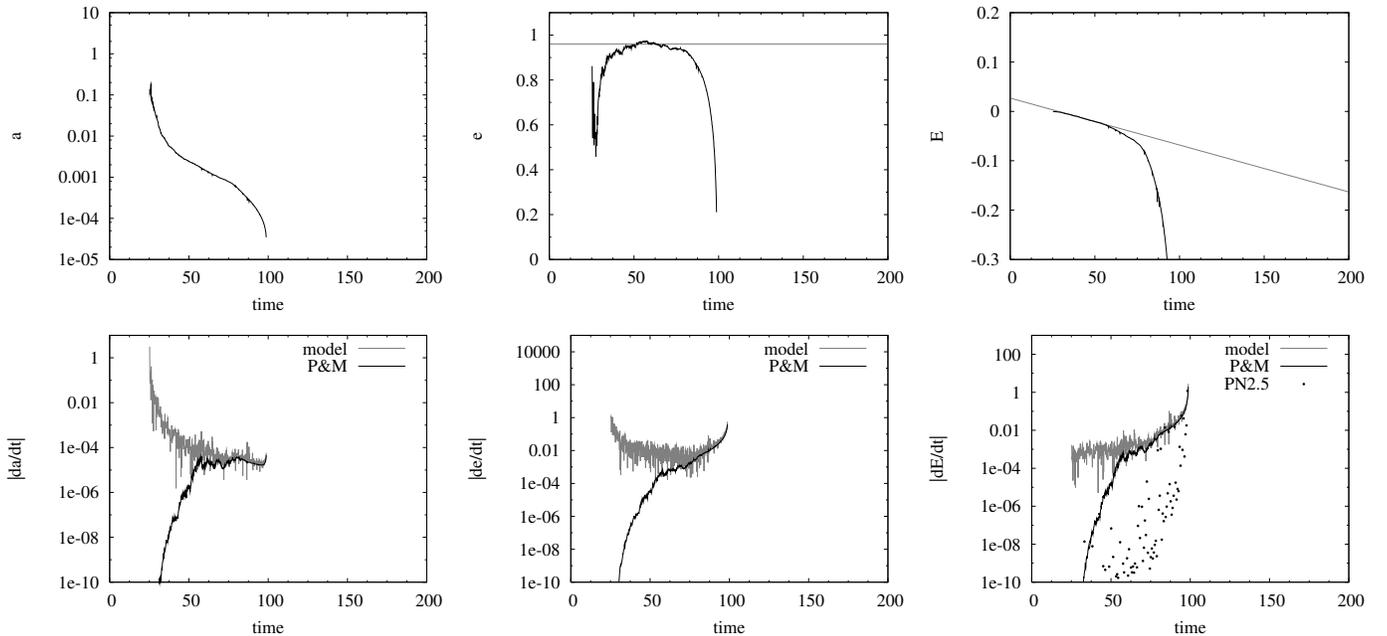}
\end{center}
\caption{Example of the time evolution a binary in the Newtonian + {\it full} $\PN$
 models with $c=447$. Same layout as in Fig.~\ref{fig5}.}
\label{fig6}
\end{figure*}

 In Fig.~\ref{figT} we plot both the calculated timescale $\tau_{\mathrm{hard}}$
 for the stellar dynamical hardening, as well as the relativistic
 coalescence timescale $t_{\mathrm{gr}}$ (see Eq.~\ref{Eq.tgr}) for orbits
 of different eccentricity as a function of their semi-major axis.
 As one can see, the two involved timescales are of the order
 of only some hundred time units or Myr, respectively, for the
 range of eccentricities found in our simulations. 
 These timescales are short enough to allow, on average, for the
 SMBH binaries to coalesce by the time the next galaxy merger would
 occur. 

 In order to test whether the purely Newtonian stellar dynamics in
 our models is sufficient enough to bring the SMBH binary into the
 regime where relativistic effects start to become important, we
 have done the following test: we first  calculate the discrete
 time-derivatives, i.e., finite differences, of the orbital
 elements and the total energy as directly computed from our
 simulations. An example is shown in Fig.~\ref{fig4} (gray curves).
 In a second step we can then use Peters \& Matthews' formalism (Eqs.~2--4)
 in order to calculate the corresponding orbit-averaged rates, which
 encode the binary's secular drift due to the emission of
 GWs at the lowest, quadrupolar, order. The resulting curves
 are also plotted in Fig.~\ref{fig4} (black curve). Finally, 
 we use the post-Newtonian expression for the (instantaneous)
 energy loss due to gravitational wave emission (Eq.~\ref{dedtpn}).

 One can see that by the time $t \approx 100$ in model units (or
 some $150$\,Myr), the two black holes have reached a separation
 that is small enough for the relativistic effects to set in and
 reach the same order of magnitude as the (Newtonian) perturbations
 from the field particles.
 This analysis demonstrates that with these rotating King models, the
 SMBH binary quickly (within a cosmological context) hardens to a point
 where it reaches the relativistic regime and the final parsec 
 problem can be overcome. In the next section, we study the
 self-consistent evolution of the (non-spinning) SMBH binaries
 in the field of stars of the galactic nuclei remnant, including
 the post-Newtonian corrections up to $2.5 \PN$ order.

\subsection{Models including Post-Newtonian Corrections} \label{SecPN}

 To test if the final parsec problem can really be overcome
 self-consistently just by stellar-dynamical effects and if
 the binary black holes in our models eventually reach the
 relativistic inspiral regime, we take the following approach:
 
 We start with models having exactly the same initial conditions
 for the set of $25$k and $50$k models described in the previous
 section, but this time we take the $\PN$ corrections for the two
 black hole particles into account. Most numerical simulations of
 similar kind which are known to us from the literature are restricted
 -- for simplicity -- to the first dissipative $\PN$ term in the expansion,
 i.e., the $2.5 \PN$ correction. The effects of the $1 \PN$ and $2 \PN$ ,
 which both are conservative and are known to result in a 
 pericenter shift, have often been neglected. However, these
 corrections are of lower order in $(v/c)$ as compared to the $2.5 \PN$
 and thus should clearly affect the binary evolution. Therefore, we
 decided to run all sets of simulations presented below with (a) only
 including the $2.5 \PN$ correction  and (b) including the full $\PN$
 corrections up to $2.5 \PN$. Since higher order corrections such
 as the $3 \PN$ term in the equations of motion are of order $c^{-6}$
 or smaller, their effect on the dynamical evolution of the BHs are
 expected to be small as well, and will not affect the main results
 presented in this work. Hence, we will loosely refer to the models
 in our set (b) as the ones with {\em full} $\PN$ corrections. We should
 note here that the $\PN$ corrections in our runs are applied permanently,
 independent of the BH separation or their relative velocities
 \citep[see for comparison][]{Aar07}. Finally, in order to enhance
 the relativistic effects we use different values of the speed
 of light $c$ (in models units).  We note that by fixing
 the values for $G$ and $c$, the ratio between the
 units of mass and length is also fixed.  Therefore, one has
 only one free parameter for setting the corresponding physical
 units in contrast to classical $N$-body simulations, which are
 scale-free in that respect.

 In Figs.~\ref{fig5} and \ref{fig6} (upper panels) we show some examples
 of the evolution of the semi-major axis $a$, the eccentricity $e$ and the
 total energy $E$ of the binary, respectively.~\footnote{
 Note that, strictly speaking, the exact expressions for the semi-major
 axis $a$, eccentricity $e$ and  energy $E$ would require $\PN$ (i.e.,
 $1 \PN$ and $2 \PN$) corrections by their own (Memmesheimer, Gopakumar
 \& Sch\"afer 2004).
 However, for a direct comparison with the Newtonian models presented in the
 previous section, we stay with the classical expressions throughout
 the rest of this work, unless stated otherwise.} The corresponding
 time derivatives, i.e., the discrete, Peters \& Mathews and $\PN$  as
 defined in the previous section are shown in the lower panels of
 these figures, respectively.
 Both simulations shown in these figures start with identical initial
 conditions and only differ by the orders of $\PN$ corrections taken
 into account.

 At early times, the dynamical evolution of the binary is qualitatively
 very similar as compared to the evolution in our purely Newtonian
 simulations (see Figs.~\ref{fig1} and \ref{fig4}): the two BH particles are rapidly
 driven towards the nuclei center by dynamical friction and the triaxial
 bar-structure in the stellar nucleus. The black holes form a pair at
 roughly $t=25$ (or some $37$\,Myr, correspondingly). Following the results
 and interpretation of \cite{BMSB06}, the initially full loss-cone around
 the black hole binary is quickly depleted and the evolution of the
 hard binary thereafter takes place in the empty loss-cone regime. Due to
 the internal rotation and the triaxial structure of the nucleus the
 loss-cone is re-populated constantly by centrophilic orbits with a rate
 which is found to be independent of the number of field particles in the
 model.

 During this phase the binary evolution is mainly dominated by just the
 Newtonian stellar-dynamical effects. The average rate of energy loss
 during this phase is again of the order of $-0.001$ as has been also
 found in the Newtonian simulations (Fig.~\ref{fig2}). Furthermore, this value is independent
 of the eccentricity with which the binary has formed and remains
 constant during the Newtonian dominated phase. That relativistic
 effects are not yet important during this phase of evolution is
 supported in the lower panels in Figs.~\ref{fig5} and ~\ref{fig6}
 in which we compare the results from the simulations with the ones
 as expected by applying Peters \& Mathews equations.

 The relativistic changes of the orbital elements become of the same
 order of the Newtonian ones after roughly
 $t\gtrsim120$ (in Fig.~\ref{fig5}) and $t\gtrsim60$ (in Fig.~\ref{fig6}).
 More generally, we find that the  time of transition from the Newtonian
 dominated regime to the one in which relativistic effects are of about
 the same order, depends strongly on the (mean) eccentricity $\bar{e}$
 of the binary's orbit.  This result is not surprising since as we have
 already stated before, the dissipation rate due to gravitational wave
 emission becomes already important at earlier times for more eccentric
 binaries.

 Finally, at about times $t \gtrsim 140$ and $70$ for Figs.~\ref{fig5}
 and \ref{fig6}, respectively, the rate of energy loss increases
 significantly due to the emission of gravitational waves and becomes
 very non-linear. The loss of orbital energy and angular momentum due
 to the $2.5 \PN$ corrections leads to the inspiral and circularization
 of the orbit, and the binary's dynamics in this phase is dominated
 by the relativistic effects, i.e., the binary almost fully completely
 decouples from the surrounding nucleus.

 It is noteworthy that these are the first systematic astrophysical
 $N$-body simulations which actually follow the evolution of the
 supermassive black holes from their unbound state, i.e. with
 order kpc separation, down to the sub-parsec scale close to relativistic
 coalescence and thus overcome the final parsec problem self-consistently.
 It seems that the origin of the latter problem partly has been the
 use of over-simplified models for the galactic nuclei, i.e., assuming
 spherically symmetric distributions without (net) angular momentum.

\section{Discussion}

%
\begin{figure}
\begin{center}
 \includegraphics[width=1.0\columnwidth]{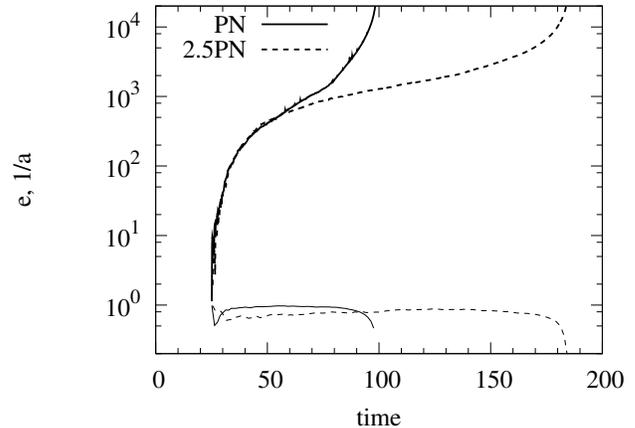}
\end{center}
\caption{Comparison of models using $2.5 \PN$ corrections only (green lines)
 and using the {\em full} $\PN$ corrections (red lines). Shown are the inverse
 of the semi-major axis (thick lines) and the eccentricity (thin lines) as
 a function of time.}
\label{rndA}
\end{figure}
 
 We have presented one of the first extensive sets of stellar-dynamical $N$-body
 simulations including full post-Newtonian corrections up to $2.5 \PN$ for the
 dominant two-body interaction between two supermassive black holes, which
 covers self-consistently all evolutionary phases from an initially
 unbound SMBH binary with separations of order $10^3$ pc down to the final
 relativistic coalescence (but see Aarseth 2003a for an early pioneering work).  Our initial conditions are still special but plausible: a ``young'' galactic
 merger remnant is represented by a flattened, rotating stellar
 nucleus with the two SMBHs at a separation of initially some $10^3$ pc.
 This is a more general
 class of initial models than used in most other papers on the subject
 \citep{mak1993,mil2001,hem2002,Aar03a,MF04,BMS05}.
 This could be seen as one possible solution of the long-standing final parsec 
 problem for black hole mergers \citep{beg1980}, as in this class
 of models we find convergence in the hardening rates at low $N$. This is the
 result of a dynamical regime where the supply of the stars into the loss-cone
 region is essentially collisionless and presumably guaranteed by a family of
 centrophilic orbits associated with this non-spherical potential
 (cf. Sections 1.1 and 5).
 
 In the following we discuss in further depth some dynamical properties of
 our models and observational consequences of our results for gravitational
 wave instruments. 

\subsection{Effect of the different $\PN$ orders}

 We have performed of order $150$ different simulations, varying the
 initial data (statistical realization, particle numbers) and different
 physical scalings (speed of light, different $\PN$ orders). In all
 models that include the $\PN$ equations of motion for the black holes
 we are able to follow the binaries' evolution down to the relativistic
 inspiral and virtually to coalescence. Based on our simulations we find
 that inclusion of the $1 \PN$ and $2 \PN$ corrections, which have been
 neglected in most earlier works (again with the notable exception of
 \cite{Aar03a}), strongly affects the dynamical evolution of the binary.
 Fig.~\ref{rndA} shows this, where we directly compare the evolution of
 $1/a$ and eccentricity as a function of time for two otherwise equal
 models with full $\PN$ and $2.5 \PN$ corrections only, respectively.
 The time of coalescence differs by almost a factor of two between these
 two cases. We find that part of the reason for this diverging behavior
 is the different eccentricity with which the binary forms.  Using full
 $\PN$ corrections rather than only $2.5 \PN$ changes the details
 of the binary's trajectory already during their first encounters.
 Consequently later on, the binary forms with different orbital elements.
 Since all close encounters in the system -- up to the one leading to the
 binary formation -- are stochastic (due to the dynamical instability of
 close few-body encounters)  we do do not find any preferred systematic
 trend in our models towards higher or lower eccentricities between the
 two types of simulations as a function of the type of $\PN$ corrections
 used.

\subsection{Orbital properties of SMBH binaries}

\begin{figure}
\begin{center}
 \includegraphics[width=1.0\columnwidth]{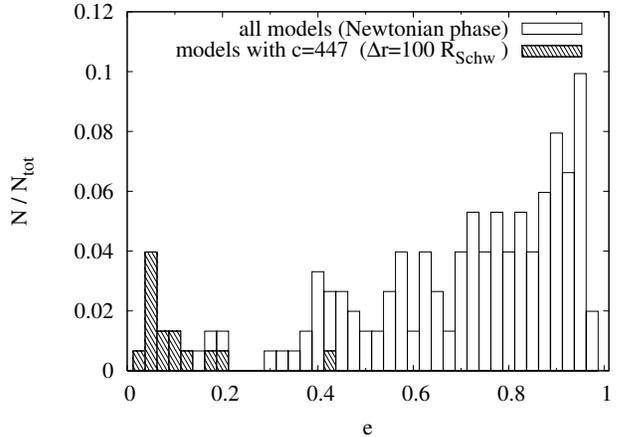}
\end{center}
\caption{Distribution of eccentricities of the binaries
 in the Newtonian regime. The white histogram shows the
 normalized distribution of eccentricities of SMBHs in
 the Newtonian evolutionary phase. The gray histograms
 shows the distribution of eccentricities at later times,
 i.e., when the binary have reached a separation of
 $100$ Schwarzschild radii. Note that at these short
 separations the majority of the orbits still has not
 yet fully circularized.}
\label{ebin}
\end{figure}

 In Fig.~\ref{ebin} we show the distribution of eccentricities
 $\bar{e}$ obtained from the sample of all our simulations.
 It clearly shows that the binaries form preferentially with
 very high eccentricities. This is favored by the transient triaxial 
 feature in our models \citep[see also,][]{BMSB06} which brings the two black holes together
 with relatively small impact parameter. Since galactic mergers lead dominantly
 to the formation of stellar bars or other triaxial configurations 
  \citep[see, e.g., ][]{Khoch06,Naab07,Joh08}, we expect that high eccentricities in
 SMBH binaries after galactic mergers are a robust phenomenon occurring
 in many real cases. Furthermore, the merger of spherical models of galactic
 nuclei along nearly-parabolic trajectories also leads to the formation of
 SMBH binaries with a similar distribution of high eccentricities
 \citep{PBBMS08}.

 It is known that during the relativistic inspiral, emission of GWs leads
 to the circularization of the binary \citep{P64}. In the same Fig.~\ref{ebin},
 we therefore also show the (shaded) histogram for the binary's eccentricities when their
 semi-major axis has fallen down to $100$ Schwarzschild radii $R_{\mathrm{BH}}$, for
 the runs with the realistic values of $c=447$. We can clearly see that, although 
 circularization has substantially reduced the binaries' eccentricity, the remaining 
 distribution at such small separation is still significantly different from zero.
 Since the orbital eccentricity of the binaries is very
 important to predict gravitational waveforms from these binaries, and the
 expected distribution of their orbital parameters strongly determines
 the complexity of the data analysis for gravitational wave instruments such
 as the planned LISA satellites \citep[e.g.,][]{Babak2008}, we are currently generalizing the present
 study to a wider range of more realistic initial conditions.
 It may
 be interesting to note that this type of statistics provides an astrophysically motivated 
 set of initial conditions for numerical relativity simulations of SMBH binary mergers.
 During the last couple of years several groups have made significant progress
 in modeling black hole mergers by the solution of the full Einstein equations and
 of numerical-relativity simulations (cf. \citet{Rezzolla2008} and citations
 $8$--$13$ therein).

%

%
\begin{figure}
\begin{center}
 \includegraphics[width=1.0\columnwidth]{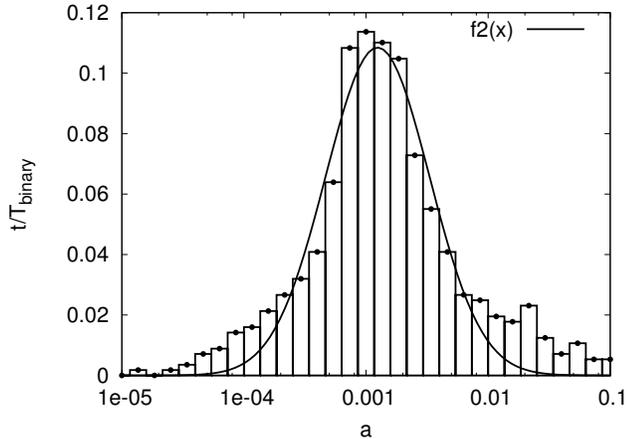}
\end{center}
\caption{Example of the time a binary spends at different
 semi-major axis intervals. The solid line shows a Gaussian
 fit to the histogram used to determine  expectation value
 of $a$ and $\sigma$.}
\label{stat2}
\end{figure}

%
\begin{figure}[ht]
\begin{center}
 \includegraphics[width=1.0\columnwidth]{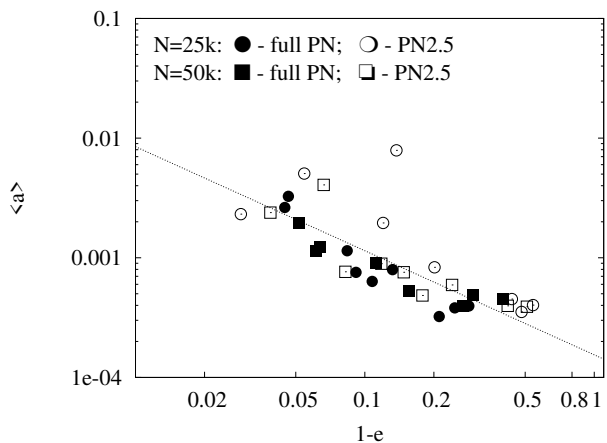}
\end{center}
\caption{Expectation value of $\langle a \rangle $ as a function
 of the orbital eccentricity for simulations with $c=447$. The
 fitted line only takes the -- more realistic -- full $\PN$ models
 into account.}
\label{statt2}
\end{figure}

%
\begin{figure}[ht]
\begin{center}
 \includegraphics[width=1.0\columnwidth]{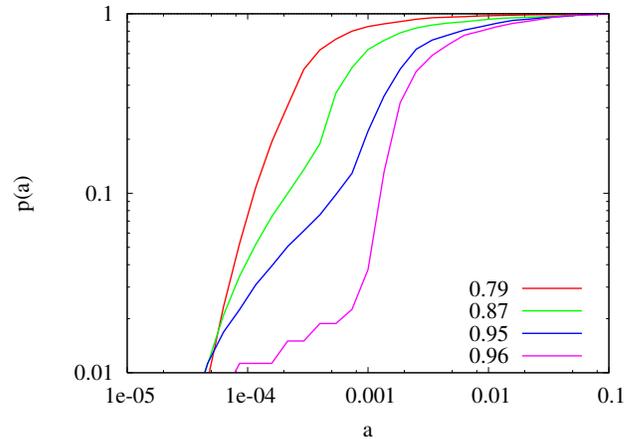}
\end{center}
\caption{Probability to find the SMBH binaries with separations
 below some given maximum semi-major axis. The different lines
 correspond to orbits with different eccentricity as given
 in the legend.}
\label{sumup2}
\end{figure}

%
\begin{figure*}[ht]
\begin{center}
 \includegraphics[width=1.0\textwidth]{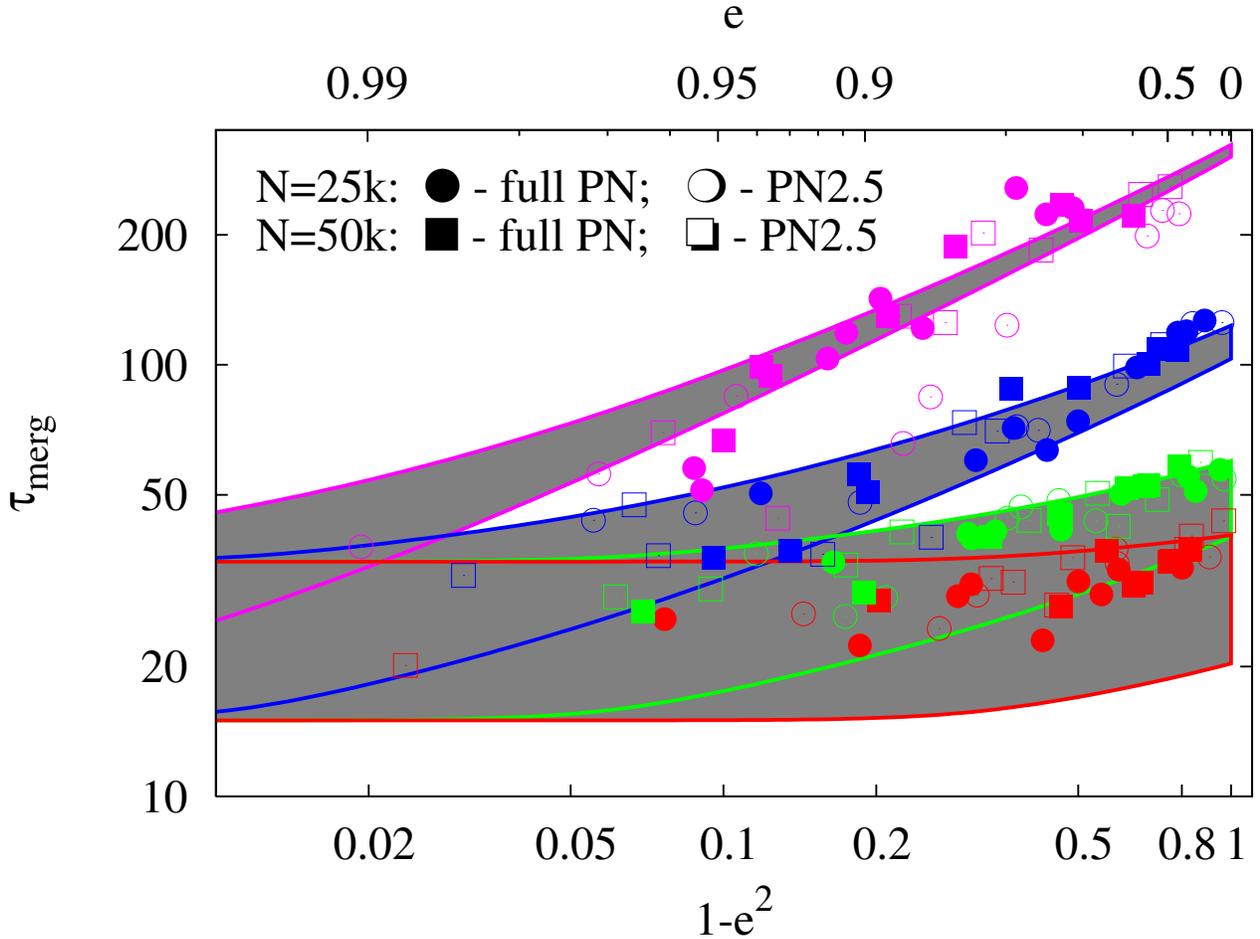}
\end{center}
\caption{Merging time $T_{\mathrm{merge}}$ of the SMBH binaries for models with different
 values of $c$ ($c=14$:black; $c=44$:red; $c=141$:green; $c=447$: blue. The
 shaded region indicate the predicted merging times using Eq.~\ref{Eq7} as
 given in the text. The upper and lower boundaries of the shaded regions
 account for different formation times $T_{\mathrm{form}}$ of the binary. Since
 the Newtonian dominated regime lasts longer for models with $c=447$, the
 eccentricity of the orbit varies stronger and results in a larger scatter
 in the plot. The are no clear differences between models with $2.5 \PN$ only
 and {\it full} $\PN$ corrections.}
\label{pgpeter}
\end{figure*}

 As we have described in Sec.~\ref{SecPN} we can distinguish
 three dynamical phases of SMBH binary evolution: (1) Newtonian
 phase, (2) mixed phase and (3) relativistic phase. It is thus
 interesting to see how much time a binary will spend the
 corresponding separations. In Fig.~\ref{stat2} we plot a
 histogram showing the time the binary spends at a given semi-major axis
 normalized to the total binary lifetime, measured from its formation
 till coalescence. The binary orbit of the model selected for
 Fig.~\ref{stat2} spends some ten per cent during its evolution at
 semi-major axis of $10^{-3}$, or $1$\,pc in physical units. We fit a
 Gaussian distribution to the binned data to provide a rough measure
 of the average value $\langle a \rangle$ and of the dispersion
 around it. 

 In Fig.~\ref{statt2} we then plot $\langle a\rangle $ as a function
 of the orbital eccentricity. While the SMBH binaries with high
 eccentricity tend to have larger $\langle a \rangle $, the fitted slope of the
 double logarithmic plot of $\langle a \rangle $ vs. $(1-e)$ is $-0.87$.
 We conclude therefore that the average pericenter $r_{\mathrm p} = \langle a \rangle (1-e)$
 of the SMBH orbit at any given average $\langle a \rangle $ has
 a remarkably small variation, i.e., is nearly constant. Therefore
 -- to first order -- the orbits of SMBH binary in our models form
 a one-parameter family. This result is relevant and interesting for
 the determination of gravitational wave backgrounds in the ultra-low
 and low-frequency areas from SMBH binaries in the universe:
 Eccentric SMBH binaries emit gravitational waves with a spectrum of
 frequencies $g(n,e)$, where $n$ denotes the higher harmonics over
 the basic mode of the circular orbit (see Appendix). The harmonic
 which leads to the maximal emission of gravitational
 waves $n_{\mathrm{max}}(e)$ is for high eccentricities ($e\ge 0.8$) completely
 determined by the frequency of motion at pericenter
 $n_{\mathrm{peri}}\propto (1-e)^{-3/2}$ \citep{Pierro2001,Amaro2008}.
 Thus for every value of average semi-major axis  $\langle a \rangle $ there
 exists a typical pericenter value and thus a typical gravitational wave
 spectrum. The expected signals of these objects (in particular highly
 eccentric ones) lie in the pulsar timing and lower LISA bands
 \citep[see for more details, including triple black holes, also][]{Enoki2007,Hoffman2007}.

 Fig.~\ref{sumup2} illustrates the cumulative probability to find the
 SMBH binary with a semi-major axis smaller than $a$. For each of our
 runs with full $\PN$ terms and realistic value of $c=447$ one curve is
 plotted in the figure. From this information
 one can deduce that SMBH binaries with high eccentricity stay longer at
 larger $a$, which is an information consistent with the previous figure. 
 The probabilities are normalized for each run separately, therefore
 these data cannot be used to determine probability distributions of
 $a$ for samples of SMBH binaries in the universe.

\subsection{Timescales for SMBH coalescence}  \label{secT}

 In Sec.~\ref{SecN} we estimated the time required for the coalescence
 of SMBH binaries in our models driven by stellar-dynamical effects
 only. On the other hand, using the pure relativistic estimate
 of the coalescence timescale $t_{\mathrm{gr}}$ (Eq.~\ref{Eq.tgr}),
 we find for typical semi-major axis values of
 $a_{\mathrm h} \approx 10^{-2} \ldots 10^{-4}$
 and eccentricities of $ e \approx 0.5 \ldots 0.99$ that the resulting timescale
 covers a range of $t_{\mathrm {gr}} \approx 10^{9} \ldots 10^{-4}$ in
 our model units. The interpretation of these numbers is that for
 small semi-major axes $a$ and high eccentricities the two SMBHs basically
 directly plunge, while for larger $a$ and moderate $e$ the relativistic
 timescale becomes unreasonable large. The evolution and hardening of
 the binary in such case would therefore mainly be driven by the Newtonian
 stellar-dynamics over some time period. Eventually, the semi-major
 axis may reach small enough values for the emission of gravitational waves
 to become efficient. The binary hardening would then equally be driven
 by both the classical Newtonian and the relativistic effects, before
 the latter eventually takes over, i.e., when the binary dynamically
 decouples from the stellar background.

 To get a more precise estimate of the time until relativistic coalescence,
 we numerically integrate the following differential equation:

\begin{equation}
 T_{\mathrm{merg}} = T_{\mathrm{form}} +  \int \left[ \left(\frac{\partial a}{\partial t}\right)_{\mathrm{hard}}
+ \left( \frac{\partial a}{\partial t} \right)_{\mathrm{P\&M}} \right] dt \, ,
\label{Eq7}
\end{equation}

 \noindent where the corresponding $da/dt$ are given by Eq.~\ref{dadtpm}
 and \ref{Eq4}, respectively. For the integration we start with
 semi-major axis separation of some $10^{-3}$ and different eccentricities.
 For the formation time of the binary we assume $T_{\mathrm{form}}=15$ and
 $35$ as a lower and upper limit, respectively.

 In Fig.~\ref{pgpeter} we plot the coalescence time as measured from
 our simulations for different values of $c$ as a function
 of eccentricity. The results are then compared to the results 
 obtained from integration of the above Eq.~\ref{Eq7}. Considering the
 different times it takes the binaries to form a bound pair
 in the simulations our results agree (within the error limits)
 with the {\em theoretical} value.

 Not surprisingly, we find that the merging time increases with
 increasing values of $c$, since the relativistic corrections
 decrease for simulations starting with the same configuration.
 However, since the hardening rate due to super-elastic scattering
 with the binary is found to be constant in our simulations, we
 expect an upper limit of $T_{\mathrm{merg}}$ of roughly $500$ time
 units.  It is important to note that in all our simulations
 the two black hole coalesce in less then  $0.5$\,Gyr after
 they have formed a bound pair! Therefore we conclude that 
 our results provide a dynamical substantiation to the picture of
 prompt SMBH coalescence advocated by \citet{Vo03} and \citet{Sesana05, Sesana07}.
 Moreover, we should also note that these times for black hole coalescences
 are of the same order as the average interval between consecutive mergers
 (cf. e.g. Fig.~$2$ in \citet{Khochfar2001}, and \citet{Khochfar2006}).
 Such timescales make it unlikely that there is enough time for a
 third black hole to interact with the binary before it has
 coalesced. If three-body interactions between SMBHs were the norm,
 then it would be quite difficult to 
 understand the notably small amount of scatter in the $M-\sigma$ relation, as
 well as the scarcity of observational evidence for off-centered nuclei. Note,
 however, that some fraction of triple interactions of SMBHs may be present
 in the universe and could cause extremely large eccentricities (e.g. through
 Kozai oscillations) and become visible through pulsar timing or even in
 the lower LISA band at comparatively large separations (orbital time
 scales) \citep{Iwasawa2006,Hoffman2007,Amaro2008,Iwasawa2008}.

\subsection{Gravitational Wave Signal}

%
\begin{figure*}
 \begin{center}
   \includegraphics[width=1.0\textwidth]{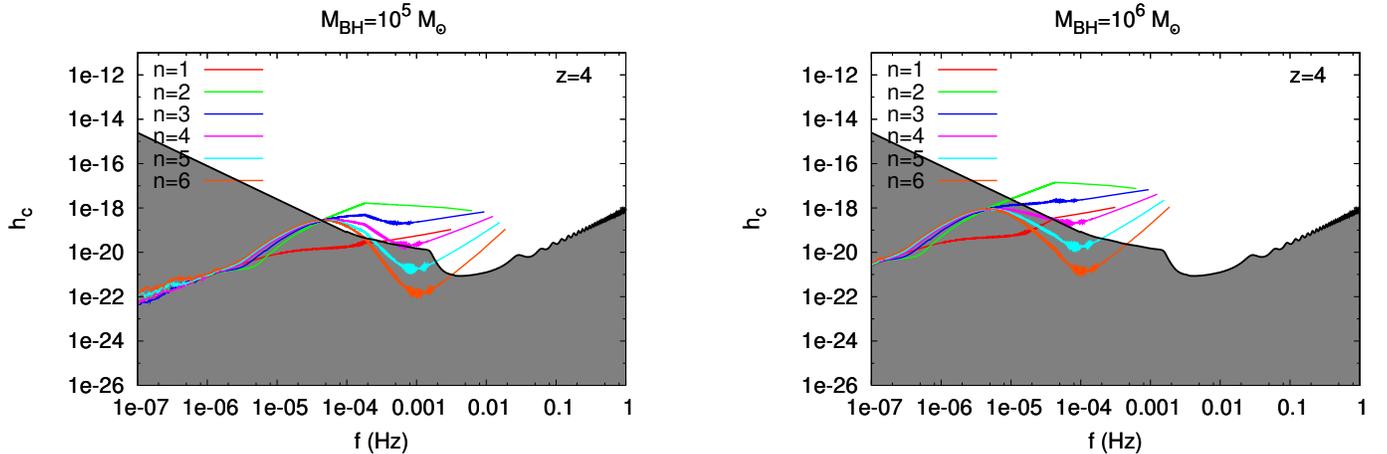}
 \end{center}
\caption{Locus of a SMBH binary from one of our simulations, for the first six
 harmonics, in the LISA sensitivity diagram during their final inspiral and
 coalescence. The black curve is the LISA sensitivity obtained from the online
 generator (see text). The dimensionless characteristic strain $h_c$ is plotted
 against the observed frequency. This case corresponds to a calculation
 including
 all $\PN$ corrections term up to the radiation reaction at $2.5\PN$ level.
 The orbital parameters adopted in this case were the full $2 \PN$ accurate
 tangential eccentricity $e_{\mathrm t}$ and semi-major axis $a$ \citep{MGS04}.
 {\bf Left panel:} the binary MBH has total mass $m_{12} = 2 \times 10^5 M_{\odot}$;
 {\bf right panel:} $m_{12} = 2 \times 10^6 M_{\odot}$; placed at redshift $z=4$.}
\label{fig13}
\end{figure*}

 Merger events of SMBH binaries are expected to be one of the
 brightest possible sources of gravitational wave emission to
 be detected by LISA \citep{Danzmann1997,Phinney2005}: The
 Laser Interferometer Space Antenna, a proposed space-borne
 gravitational wave observatory, scheduled for launch in 2018+,
 is most sensitive to gravitational waves in the low-frequency
 regime ($10^{-4}$ -- $0.1$\,Hz); for SMBHs of more than
 about $10^7 {\mathrm M}_\odot$ LISA will preferentially detect the final
 merger and ringdown phases \citep{Babak2008}, while for smaller masses
 earlier evolutionary phases of SMBH evolution will become detectable
 in the LISA band, in particular for eccentric SMBHs (see discussion
 and references in  Sect.~\ref{secT}).

 In most of the present literature the strain
 amplitude of the gravitational wave is usually estimated under
 the assumption of circular orbits. This has been motivated 
 by the idea that massive binary systems are expected to
 have been completely circularized due to the emission of
 gravitational waves \citep[see, e.g., ][]{P64, Sesana05} by
 the time they enter the LISA frequency band.
 In this section we will use our results to test this assumption.
 The details on how to extract the information on the gravitational
 wave strain amplitude from our numerical simulations calculation
 are given in the Appendix.

 The space-based gravitational wave detector LISA will be sensitive
 to signals of inspiralling compact objects with up to the maximum
 total mass $M_{12} \approx 7.5 \times 10^6$ M$_{\odot}$, where
 $M_{12}$ is the total mass of the binary. Due to the involved
 low frequencies of the signal, such sources are out of reach
 of current ground-based detectors. The physical units adopted in
 Sec.~\ref{SecInit} translate into black hole masses of about
 $10^9$ M$_{\odot}$ in our simulations. Black hole binaries in
 this mass regime, however, will inspiral with frequencies even
 lower than those of the LISA band and thus would not be detectable
 by LISA before their actual coalescence.

 In order to follow the black hole binary inspiral up to frequencies
 in range of interest for LISA, i.e, $f \geq 10^{-4}$ Hz, we first
 need to rescale our models to binary masses of order of $10^5$ or
 $10^6$ M$_{\odot}$. As mentioned earlier, the speed of light in model
 units scales as $c \propto (R/M)^{1/2}$, where $R$ and $M$ are
 the units for length-scale and mass, respectively. Since $c$
 is already given with a fixed value from our simulations, by changing
 our unit of mass by factors $10^{-3}-10^{-4}$, the length-scale
 automatically changes to $0.1$ or $1$\,pc, respectively. Finally,
 following the results of \cite{Sesana07}, we {\it place} the
 system at the redshift $z=4$, where LISA's detection rate for objects
 of that mass range is expected to be significant.

 Before actually calculating the strain amplitude of the gravitational
 wave signal,  we increase the time-resolution of our output data for
 the black hole's trajectory during the late phases of inspiral. This
 is done by evolving the binary in isolation, starting with initial
 conditions (in the binaries center of mass frame) as given from the
 large-scale simulations at a stage when it is already decoupled from
 the surrounding stellar system, i.e., when the binary evolution is
 dominated by relativistic dynamics (compare Fig.~\ref{fig6}). We
 convince ourselves that the binary is really decoupled from the 
 stellar system by comparing the evolution of its orbital elements
 to the one of the "low" time-resolution. In fact, during the
 late stages of inspiral the perturbations from stars vary over
 timescales much longer than both the orbital period and the time
 for coalescence (at that moment) which allows us to safely ignore
 them.

 Using the information of the combined time-series of the binary
 evolution we then calculate the orbital elements in the $\PN$
 generalization \citep{MGS04}. These allow us to calculate the
 different modes of the characteristic strain amplitude $h_{c,n}$
 using Eq.~\ref{eq-strainamp} from the Appendix.

 In Fig.~\ref{fig13} we show $h_{c,n}$ for the first
 six harmonics of the signal generated by the inspiral in one
 of our $50$k models for (rescaled) binary masses of 
 $2 \times 10^{5}$ M$_{\sun}$ (left panel) and $ 2 \times 10^{6}$ M$_{\sun}$ (right
 panel).  The LISA sensitivity curve $S_h(f)$ displayed in the figures
 was generated by the Online Sensitivity Curve Generator with default
 settings, corresponding to a three year observation period
 \citep{Larson07}. We confirm that the BH binaries in the mass range
 $10^5-10^6$ M$_{\sun}$ indeed enter the LISA frequency band during
 the relativistic inspiral in our simulations. The height of $h_{c,n}$
 above the sensitivity curve in Fig.~\ref{fig13} provides an
 approximate indication of the signal-to-noise ratio (SNR; see
 also Appendix) and is found to be high enough for detection by LISA.  

 As a consequence of the very high initial eccentricity with which
 the MBH binaries form in our simulations (compare Fig.~\ref{ebin}),
 they reach the LISA band with an eccentricity which may be significantly
 different from zero -- the exact distribution depending on the 
 redshift $z$ of the sources (Preto et al., in prep.). As a result,
 there will be a significant contribution to the measured power from
 higher harmonics $f_n = n\,f_{\mathrm{orb}}/(1+z)$, with $n \geq 2$ (note
 that $n=2$ for a circular orbit). This can be seen from the plots
 shown here, where several harmonics contribute significantly to
 the GW signal and are well above the threshold for detection.
 Furthermore, we find that, as the binary chirps in the LISA band, it
 circularizes quite rapidly with the consequence that the $n=2$ harmonic
 always becomes dominant at the last stages of the inspiral.

 These findings may have an important impact on the accurate computation
 of the inspiral waveforms as well as potentially leading to an
 increased upper limit of the range of detectable masses by LISA.
 Based on our results we suggest that orbits with non-vanishing
 eccentricities should indeed seriously be considered for the
 LISA data analysis. Further details and discussion of the
 consequences for the LISA detection of SMBH binary inspirals
 are beyond the scope of this work and will be published elsewhere
 (Preto et al., in prep.).

\section{Conclusions}

 We present the first $N$-body models which self-con\-sis\-tently follow
 the evolution of binary supermassive black holes in merged galaxies
 from kiloparsec separations down to gravitational-wave-induced coalescence,
 and in which the early evolution of the binary is driven by collisionless
 loss-cone repopulation, allowing the results to robustly be scaled to real galaxies.
 Our simulations include post-Newtonian corrections to the equations of motion
 of the SMBH binary up to order $2.5 \PN$; we show that inclusion of the
 energy-conserving $1 \PN$ and $2 \PN$ terms is also crucial for obtaining the
 correct time dependence of the binary orbital parameters. We identify
 evolutionary phases in which the binary is still evolving due to stellar
 encounters but in which relativistic corrections to its two-body motion
 are also important, thereby showing that the consideration of all $\PN$ orders
 {\em in} the $N$-body simulations is necessary for an accurate prediction
 of its orbital elements evolution. The SMBH binaries in our simulations
 often form with large eccentricities, and these high eccentricities are
 maintained during the Newtonian phases of the evolution. We show that the
 gravitational wave signal measured by an interferometer like LISA would contain significant
 contributions from high-order harmonics from GWs emitted from binaries at
 cosmological distances,  and that the nature of the signal can depend strongly
 on the dynamical history of the binary prior to its entering the gravitational
 wave dominated regime.

 We have shown that supermassive black hole binaries in galactic
 nuclei can overcome the stalling barrier and will reach the
 relativistic coalescence phase in a timescale shorter than the
 age of the universe. A gravitational wave signal expected for
 the LISA satellite from these SMBH binaries is expected, in
 particular due to the high eccentricity of the SMBH binary when
 entering the relativistic coalescence phase. We present for the
 first time a comprehensive set of models which cover self-consistently
 the transition from the Newtonian dynamics, dynamical friction phase
 (with yet unbound SMBH binary) to the situation when relativistic,
 post-Newtonian corrections start to influence the relative SMBH
 motion. After the shrinking time scale becomes very short the binary
 decouples from the rest of the galactic nucleus and can be treated
 as a relativistic two-body problem.
 We follow this evolution formally to the coalescence of the two black
 holes using $\PN$ terms of up to order $2.5 \PN$ and determine the
 gravitational wave emission in different modes relative to the
 LISA sensitivity curve. We find that the orbital parameters of
 a SMBH binary, when entering the LISA band, depends on the
 previous dynamical history - in particular there exists a phase
 where the SMBH binary is still partially coupled to the stellar
 environment via three-body encounters, but relativistic
 corrections to its two-body motion already play a role. 

 Since purely Newtonian models of SMBH binaries in rotating galactic nuclei
 \citep{BMSB06} have shown a {\color{black} quasi-$N$-independence of the
 stellar-dynamical driven hardening rates, we conclude that the results
 obtained in this paper regarding the time required until relativistic
 merger holds for galactic nuclei with realistic parameters.}
 In this work we limited ourselves to models with particle
 numbers up to $50$k only, however, due to the independence
 of the results in the Newtonian phase from $N$ in the current
 work and in \citet{BMSB06}, we do not expect any significant
 changes in our main results for models with $N>50$k.

 It should be noted, however, that our results show a strong
 dependence on initial conditions, and that our initial model
 is a very simple approximation to a post-merger galactic nucleus.
 Therefore our conclusion from this work can only be that it
 is possible to reach relativistic coalescence in a reasonable
 time ($10^8$ years).
 Any prediction of event rates for LISA would, however, require a more
 careful estimate of the distribution of parameters for a realistic
 set of mergers, e.g. by taking data from semi-analytic merger
 trees, similar to \cite{Sesana05}. This is the subject of presently
 ongoing work.

 We have also examined the effect of using $2.5 \PN$ alone or {\em full} $\PN$
 corrections. Simple two-body $\PN$ experiments show significant
 differences of the BHs trajectories, and we find generally a
 dependence of the merging time in the two cases. Different orbital
 parameters of the SMBH binary in the final phase would have a major
 impact on the waveforms of the GWs. However, the computation of the latter
 also
 requires simulations with higher $\PN$ corrections, at least
 up to order ${\cal O}({c}^{-6})$, i.e., $3 \PN$ in the equations
 of motion. This is beyond the scope of the current paper.

 Results on whether our simulated SMBH binary will fall into the LISA
 frequency band and at which frequencies will be discussed in a 
 cosmological context in a forthcoming paper.


\acknowledgments

 We would like to thank Achamveedu Gopakumar, Gerhard Sch\"afer and
 Sverre Aarseth for enlightening and fruitful discussions on various
 aspects of the post-Newtonian dynamics, and Pau Amaro-Seoane and
 Gabor Kupi for discussions about the numerical implementation.
 {\color{black} We are particularly grateful to Sverre Aarseth for
 helping to improve the manuscript.} {\color{black} We also thank
 the referee for his/her constructive comments.}
 Financial support for this work was provided by project {\sc 'GRACE'}
 I/80\,041-043 of the Volkswagen Foundation and by the Ministry of Science,
 Research and the Arts of Baden-W\"urttemberg (Az: 823.219-439/30 and /36).
 We also acknowledge funding by DLR (Deutsches Zentrum f\"ur Luft- und
 Raumfahrt) and Astrogrid-D through the German Ministry of Education and
 Research (BMBF). This project is also partly funded by the German Science
 Foundation (DFG) under SFB~439 (sub-project B11) \textsl{``Galaxies in
 the Young Universe''} at the University of Heidelberg. Furthermore we
 acknowledge a computing time grant obtained from the DEISA project with
 FZ J\"ulich. PB thanks for the special support of his work by the
 Ukrainian National Academy of Sciences under the  Main Astronomical
 Observatory 'GRAPE/GRID' computing cluster  project.



\appendix

\section{Appendix material}

 Any gravitational wave carries energy: The total energy carried by a
 wave is $E \sim N(f) \ h^2$, where $N(f)$ is the number of cycles the
 wave spends on a frequency interval $\Delta f \sim f$ around the
 frequency $f$. It is customary to define a {\it characteristic strain}
 $h_{\mathrm c}$ corresponding to an observation with duration
 $\tau \gtrsim N(f)/f$ by $h_{\mathrm c}(f)=\sqrt{N(f)} \ h(f)$.
 In this case, the signal is not monochromatic and we observe its chirp 
 as it shifts to higher frequency during the observation. However, in
 case the observation time $\tau \lesssim N(f)/f$, the signal is
 approximately monochromatic and its amplitude is essentially limited
 by the observing time rather than by the intrinsic properties of the
 system and, as a result, $h_{\mathrm c} = \sqrt{\tau f} \ h(f)$ \citep{KT87}.

 In the relativistic inspiral phase, the back-reaction from the GW
 emission (energy balance argument) dominates the binary's orbital
 decay \citep{Bla06}. The number of cycles around a given frequency
 $f$ can be estimated to be

\begin{equation}
 N(f) \sim {\dot f_{\mathrm r}}/{f_{\mathrm r}^2} =
 \frac{5 {c}^5}{96 \pi^{8/3}} {G}^{-5/3} \mathcal{M}^{-5/3} f_{\mathrm r}^{-5/3} ,
\end{equation}

 \noindent
 where the orbit is described in terms of the (instantaneous) Kepler elements,
 the binary's chirp mass
 is $\mathcal{M} = m_1^{3/5} m_2^{3/5}/(m_1+m_2)^{1/5}$, and $f_{\mathrm r}$
 is the orbital frequency in the source's rest frame. Following
 \cite{PM63}, the orbital frequency shifts at a rate given by

\begin{equation}
\dot f_{\mathrm r} = \frac{df_{\mathrm r}}{da} \frac{da}{dt} = \frac{96 \pi^{8/3}}{5 c^5} G^{5/3}
\mathcal{M}^{5/3} f_r^{11/3}.
\label{eq-freqshiftrate}
\end{equation}

 \noindent
 The solution $f_{\mathrm r}$ to Eq.~\ref{eq-freqshiftrate}
 blows up in a finite time $\tau_{\mathrm{coal}}$, which is used to
 denote the coalescence time.  For a given frequency $f_{\mathrm r}$ the
 time till coalescence can by calculated according to the following
 expression:

\begin{equation}
\tau_{{\mathrm{coal}}} \approx \frac{N(f_{\mathrm r})}{f_{\mathrm r}} \approx
 \frac{5 {c}^5}{96\pi^{8/3}} {G}^{-5/3} \mathcal{M}^{-5/3}
 f_{\mathrm r}^{-8/3} .
\end{equation}

 Assuming a Keplerian orbital parametrization, the strain amplitude
 for the $n^{\mathrm{th}}$ harmonic (after sky averaging) is given by

\begin{equation}
h_{n}(f) = \sqrt{h_{+}^2 + h_{\times}^2} = \frac{37 \pi^{2/3}}{16}
 \frac{G^{5/3} \mathcal{M}^{5/3}}{r(z) {c}^4} f_{\mathrm r}^{2/3}
 \sqrt{g(n,e)}.
\end{equation}

Note that $r(z)$ is the comoving distance (as a function of cosmological
redshift $z$) to the source and $g(n,e)$ represents the normalized relative
power spectrum in the $n^{\mathrm{th}}$ harmonic of the GW signal \citep{PM63}:

\begin{eqnarray}
g(n,e) = \frac{n^4}{34} \left\{ \left[J_{n-2}(ne)-2eJ_{n-1}(ne)+\frac{2}{n}
J_n(ne)+2eJ_{n+1}(ne)
                                                                -J_{n+2}(ne)\right]^2
\right. \nonumber \\
\left.      + (1-e^2) \left[J_{n-2}(ne)-2J_n(ne)+J_{n+2}(ne) \right]^2 +
\frac{4}{3n^2} J_n^2(ne)
\right\},
\end{eqnarray}

 \noindent
 where the $J_{n}$ are Bessel functions of the first kind. The characteristic
 strain $h_{c,n}$ can therefore be written as

\begin{eqnarray}
\label{eq-strainamp}
 h_{c,n}(f) & = &  \sqrt{N} \ h \sim \frac{37}{64 \pi^{2/3}} \sqrt{5/6} \frac{{G}^{5/6}
      \mathcal{M}^{5/6}}{c^{3/2} R(z)} \ \sqrt{g(n,e)} \ f_{\mathrm{obs}}^{-1/6},
                        \hspace{1.8em} N \lesssim f_{\mathrm{obs}} \,\tau   \nonumber  \\
                     \\
h_{c,n}(f) & = & \sqrt{f \tau} \ h \sim \frac{37 \pi^{2/3}}{16}
\frac{G^{5/3}\mathcal{M}^{5/3}}{c^4 R(z)}
      \sqrt{g(n,e)} \ \sqrt{\tau} \ f_{\mathrm{obs}}^{7/6},  \hspace{3em} N \gtrsim
 f_{\mathrm{obs}}\,\tau.   \nonumber
\end{eqnarray}
 
 Note that the frequency seen in the detector is given by
 $f_{\mathrm{obs}}=f_{\mathrm r}/(1+z)$, where $f_{\mathrm{obs}}$ is the
 frequency measured in the binary's reference frame. The knee
 frequency separates the two regimes

\begin{equation}
f_{\mathrm{knee}} = N(f) f_{\mathrm{obs}} \approx \frac{1}{\pi} \left( \frac{5}{96 \tau}
\right)^{3/8}
              \frac{{c}^{15/8}}{{G}^{5/8}\mathcal{M}^{5/8}}\, \left( 1+z \right)^{-1}.
\end{equation}

\noindent
 Note that the knee frequency decreases if the observation time $\tau$
 increases.

 The GW signal is said to come {\it into band} when
 $h_{c,n}(f) \geq \langle f \ S_h(f) \rangle^{1/2}$,
 where $S_h(f)$ is the instrumental strain noise spectrum (in Hz$^{-1}$)
 and the brackets denote sky-averaging. The LISA sensitivity curve
 $S_{\mathrm h}(f)$ displayed in the figures was generated by the Online
 Sensitivity Curve Generator with default settings, corresponding to
 a three year observation period \citep{Larson07}. The height of $h_c$
 above the sensitivity curve provides an approximate indication of the
 signal-to-noise ratio (SNR); however, the integrated SNR should be
 computed as

\begin{equation}
\left( \mathrm{SNR} \right)^2 = \int_{f_{{\mathrm{min}}}}^{f_{{\mathrm{max}}}} \frac{d f}{f}
                                \ \frac{h_c^2(f)}{\langle f \ S_h(f) \rangle}.
\end{equation}


\begin{thebibliography}{}

\bibitem[Aarseth(1985)]{Aar85}
Aarseth, S.J.\ 1985, in Direct Methods for $N$-body Simulations, 
 ed. J.U. Brackbill, \& B.I. Cohen (Academic Press), p.~337

\bibitem[Aarseth(1999)]{Aar99}
Aarseth, S.J.\ 1999, \pasp, 111, 1333

\bibitem[Aarseth(2003a)]{Aar03a}
Aarseth, S.J.\ 2003a, \apss, 285, 367

\bibitem[Aarseth(2003b)]{Aar03b}
Aarseth, S.J.\ 2003b, Grav\-i\-ta\-tional $N$-body Sim\-u\-la\-tions
 (Cam\-bridge, UK: Cam\-bridge Uni\-versity Press)

\bibitem[Aarseth(2007)]{Aar07}
Aarseth, S.J.\ 2007, \mnras, 378, 285

\bibitem[Amaro-Seoane et al.(2008)]{Amaro2008}
Amaro-Seoane, P., Benacquista, M., Hoffman, L., Eichhorn, C., Spurzem, R., \&
 Makino, J., submitted to \mnras

\bibitem[Andrade, Blanchet \& Faye(2001)]{ABF01}
Andrade, V.C., Blanchet, L., \& Faye, G.\ 2001, Class. Quantum Grav., 18, 753

\bibitem[Babak et al.(2008)]{Babak2008}
Babak, S., Hannam, M., Husa, S., \& Schutz, B.\ 2008, arXiv:0806.1591

\bibitem[Begelman et al.(1980)]{beg1980}
Begelman M.C., Blandford R.D., \& Rees M.J.\ 1980, \nat, 287, 307

\bibitem[Berentzen et al.(2008)]{BPBMS08}
Berentzen, I., Preto, M., Merritt, D., Berczik, P., \& Spurzem, R.\ 2008, Astron. Nachr., 329, 904

\bibitem[Berczik et al.(2005)]{BMS05}
Berczik, P., Merritt, D., \& Spurzem, R.\ 2005, \apj, 633, 680

\bibitem[Berczik et al.(2006)]{BMSB06}
Berczik, P., Merritt, D., Spurzem, R., \& Bischoff, H.-P.\ 2006, \apj, 642, L21

\bibitem[Berentzen et al.(2008)]{BPBS}
Berentzen, I., Preto, M., Berczik, P., \& Spurzem, R.\ 2008, Astron. Nachr., 329, 904

\bibitem[Bianchi et al.(2008)]{bia08}
Bianchi, S., Chiaberge, M., Piconcelli, E., Guainazzi, M., \& Matt, G.\ 2008, \mnras, 386, 105

\bibitem[Blanchet(2006)]{Bla06}
Blanchet, L., 2006, in Living Rev. Relativity 9, 4. URL (cited on June 2007):
 http://www.livingreviews.org/lrr-2006-4

\bibitem[Danzmann(1997)]{Danzmann1997}
Danzmann, K.\ 1997, Class. Quantum Grav., 14, 1399

\bibitem[Dotti et al.(2007)]{Dotti07}
Dotti, M., Colpi, M., Haardt, F., \& Mayer, L.\ 2007, \mnras, 379, 956

\bibitem[Enoki \& Nagashima(2007)]{Enoki2007}
Enoki, M., \& Nagashima, M.\ 2007, Progress of Theoretical Physics, 117, 241 

\bibitem[Ernst et al.(2007)]{EGFJS07}
Ernst, A., Glaschke, P., Fiestas, J., Just, A., \& Spurzem, R.\ 2007, \mnras, 377, 465

\bibitem[Escala et al.(2004)]{Escala04}
Escala, A., Larson, R.~B., Coppi, P.~S., \& Mardones, D.\ 2004, \apj, 607, 765

\bibitem[Escala et al.(2005)]{Escala05}
Escala, A., Larson, R.~B., Coppi, P.~S., \& Mardones, D.\ 2005, \apj, 630, 152

\bibitem[Ferrarese \& Ford(2005)]{FF05}
Ferrarese, L., \& Ford, H.\ 2005, Space Sci. Rev., 116, 523

\bibitem[Fukushige et al.(2005)]{FMK05}
Fukushige, T., Makino, J., \& Kawai, A.\ 2005, \pasj, 57, 1009

\bibitem[Gerhard \& Binney(1985)]{GB85}
Gerhard O.E. \& Binney J.\ 1985, \mnras, 216, 467

\bibitem[Harfst et al.(2007)]{HGMS07}
Harfst, S., Gualandris, A., Merritt, D., Spurzem, R., Portegies Zwart, S.,
 \& Berczik, P. 2007, New A, 12, 357

\bibitem[Hemsendorf, Sigurdsson \& Spurzem(2002)]{hem2002}
Hemsendorf M., Sigurdsson S., \& Spurzem R.\ 2002, \apj, 581, 1256

\bibitem[Hoffman \& Loeb(2007)]{Hoffman2007}
Hoffman, L., \& Loeb, A.\ 2007, \mnras, 377, 957

\bibitem[Iwasawa, Funato \& Makino(2006)]{Iwasawa2006}
Iwasawa, M., Funato, Y., \& Makino, J.\ 2006, \apj, 651, 1059

\bibitem[Iwasawa, Funato \& Makino(2008)]{Iwasawa2008}
Iwasawa, M., Funato, Y., \& Makino, J.\ 2008, arXiv:0801.0859

\bibitem[Jesseit et al.(2007)]{Naab07}
Jesseit, R., Naab, T., Peletier, R. F., \& Burkert, A.\ 2007, \mnras, 376, 997 

\bibitem[Johansson, Naab \& Burkert(2008)]{Joh08}
Johansson, P. H., Naab, T., \& Burkert, A., arXiv:0802.0210 

\bibitem[Kazantzidis et al.(2005)]{Kazant05}
Kazantzidis, S., et al.\ 2005, \apjl, 623, L67

\bibitem[Khochfar \& Burkert(2001)]{Khochfar2001}
Khochfar, S., \& Burkert, A.\ 2001, \apj, 561, 517

\bibitem[Khochfar \& Burkert(2006)]{Khochfar2006}
Khochfar, S., \& Burkert, A.\ 2006, \aap, 445, 403

\bibitem[King(1966)]{King66}
King, I.R.\ 1966, \aj, 71, 276

\bibitem[Komossa et al.(2003)]{kom2003} Komossa
S., Burwitz V., Hasinger G., Predehl P., Kaastra J.S., \& Ikebe, Y.\ 2003, \apj, 582, L15

\bibitem[Kupi et al.(2006)]{KASS06}
Kupi, G., Amaro-Seoane, P., \& Spurzem, R.\ 2006, \mnras, 371, L45

\bibitem[Larson(2007)]{Larson07}
Larson, S.L., 2007, in A modern astrophysical laboratory, in LISC, the LISA International
 Science Community \\ http://www.lisa-science.org/resources/introductory-lisa-material, 2007.

\bibitem[Libeskind et al.(2006)]{Libes06}
Libeskind, N.~I., Cole, S., Frenk, C.~S., \& Helly, J.~C.\ 2006, \mnras, 368, 1381

\bibitem[L\"ockmann \& Baumgardt(2008)]{LB08}
L\"ockmann, U., \& Baumgardt, H.\ 2008, MNRAS, 384, L323

\bibitem[Longaretti \& Lagoute(1996)]{LL96}
Longaretti, P.-Y., \& Lagoute, C.\ 1996, \aap, 308, 453

\bibitem[Madau et al.(2004)]{Madau04}
Madau, P., Rees, M.~J., Volonteri, M., Haardt, F., \& Oh, S.~P.\ 2004, \apj, 604, 484

\bibitem[Makino(1997)]{Mak07}
Makino, J.\ 1997, \apj, 478, 58

\bibitem[Makino \& Aarseth(1992)]{MA92}
Makino, J., \& Aarseth, S.J.\ 1992, \pasj, 44, 141

\bibitem[Makino et al.(1993)]{mak1993}
Makino J., Fukushige T., Okumura S.K., \& Ebisuzaki T.\ 1993, \pasj, 45, 303

\bibitem[Makino \& Funato(2004)]{MF04}
Makino, J., \& Funato, Y.\ 2004, \apj, 602, 93

\bibitem[Mayer et al.(2007)]{Mayer07}
Mayer, L., Kazantzidis, S., Madau, P., Colpi, M., Quinn, T., \& Wadsley, J.\ 2007, Science, 316, 1874

\bibitem[Memmesheimer et al.(2004)]{MGS04}
Memmesheimer, R.M., Gopakumar, A., \& Sch\"{a}fer, G.\ 2004, \prd, 70, 104011

\bibitem[Merritt(2006)]{Merritt06}
Merritt, D.\ 2006, \apj, 648, 976
\bibitem[Merritt \& Milosavljevi{\'c}(2005)]{MM05}
Merritt, D., Milosavljevi{\'c} M.\ 2005, Living Rev. Relativity 8, 8. URL\footnote{http://www.livingreviews.org/lrr-2005-8}

\bibitem[Merritt \& Poon(2004)]{MPoon04}
Merritt D., \& Poon M.Y.\ 2004, \apj, 606, 788

\bibitem[Merritt, Mikkola \& Szell(2007)]{MMS07}
Merritt D., Mikkola S, \& Szell A.\ 2007, \apj, 671, 53

\bibitem[Milosavljevi\'c \& Merritt(2001)]{mil2001}
Milosavljevi{\'c} M., \& Merritt D.\ 2001, \apj, 563, 34

\bibitem[Naab, Khochfar, \& Burkert(2006)]{Khoch06}
Naab, T., Khochfar, S., \& Burkert, A.\ 2006, \apj, 636, L81

\bibitem[Peters(1964)]{P64}
Peters, P.C.\ 1964, \prb, 136, 1224

\bibitem[Peters \& Mathews(1963)]{PM63}
Peters, P.C., \& Mathews, J.\ 1963, Phys. Rev., 131, 435

\bibitem[Phinney(2005)]{Phinney2005}
Phinney, S.\ 2005, in General Relativity and Gravitation, ed. P. Florides, 
 B. Nolan, \& A. Ottewill, Proceedings of the 17th International Conference
 held at RDS Convention Centre, Dublin, Ireland, July 18-23, 2004. QC173.6.I57 2004;
 ISBN 981-256-424-1. (World Scientific Publishing Co., Ltd., London, England),  p.~118

\bibitem[Pierro et al.(2001)]{Pierro2001}
Pierro, V., Pinto, I.~M., Spallicci, A.~D., Laserra, E., \& Recano, F.\ 2001, \mnras, 325, 358

\bibitem[Preto et al.,(submitted)]{PBBMS08}
Preto, M., Berentzen, I., Berczik, P., Merritt, D., \& Spurzem, R.\ 2008,
submitted to Journal of Physics. arXiv:0811.3501

\bibitem[Rezzolla et al.(2008)]{Rezzolla2008}
Rezzolla, L., Barausse, E., Dorband, E.~N., Pollney, D., Reisswig, C., Seiler, J.,
 \& Husa, S.\ 2008, \prd, 78, 044002

\bibitem[Rodriguez et al.(2006)]{Rod06}
Rodriguez, C., Taylor, G.~B., Zavala, R.~T., Peck, A.~B., Pollack, L.~K.,
\& Romani, R.~W.\ 2006, \apj, 646, 49

\bibitem[Schutz(1985)]{Sch85}
Schutz, B.F. 1985, A First Course in General Relativity.  Cambridge University Press, 1985

\bibitem[Sesana et al.(2007)]{Sesana07}
Sesana, A., Volonteri, M., Haardt, F., 2007, \mnras, 377, 1711

\bibitem[Sesana et al.(2005)]{Sesana05}
Sesana, A., Haardt, F., Madau, P., \& Volonteri, M.\ 2005, \apj, 623, 23

\bibitem[Thorne(1987)]{KT87}
Thorne, K.S., In: Three Hundred Years of Gravitation, 1987, Cambridge University Press, Cambridge

\bibitem[Valtonen(2007)]{Val07}
Valtonen, M.~J.\ 2007, \apj, 659, 1074

\bibitem[Volonteri et al.(2003)]{Vo03}
Volonteri, M., Haardt, F., \& Madau, P.\ 2003, \mnras, 582, 559

\bibitem[Yu(2002)]{Yu02}
Yu, Q.\ 2002, \mnras, 331, 935
\end{thebibliography}
\end{document}